\documentclass[twocolumn]{aastex701}

\usepackage{graphicx}
\usepackage{amsmath}
\usepackage{array}
\usepackage[T1]{fontenc}
\usepackage{afterpage}

\usepackage{ulem}
\DeclareRobustCommand{\del}{\bgroup\markoverwith{\textcolor{magenta}{\rule[.5ex]{2pt}{0.4pt}}}\ULon}

\begin{document}

\title{Long-Term Evolution of Close-in Sub-Neptunes and Outer Planetary Embryos: Atmospheric Mass Loss and Origin of Planets Inside and Outside the Radius Gap}

\author[0000-0001-5264-1924,sname='He']{Yaxing He}
\affiliation{State Key Laboratory of Dark Matter Physics, Tsung-Dao Lee Institute, Shanghai Jiao Tong University, 1 Lisuo Road, Shanghai 201210, China}
\email{yaxinghe@sjtu.edu.cn}  

\author[0000-0002-8300-7990, sname='Ogihara']{Masahiro Ogihara}
\affiliation{State Key Laboratory of Dark Matter Physics, Tsung-Dao Lee Institute, Shanghai Jiao Tong University, 1 Lisuo Road, Shanghai 201210, China}
\affiliation{School of Physics and Astronomy, Shanghai Jiao Tong University, 800 Dongchuan Road, Shanghai 200240, China}
\email{ogihara@sjtu.edu.cn}
% \correspondingauthor{Masahiro Ogihara}

\author[0000-0001-6870-3114, sname='Guo']{Kangrou Guo}
\affiliation{State Key Laboratory of Dark Matter Physics, Tsung-Dao Lee Institute, Shanghai Jiao Tong University, 1 Lisuo Road, Shanghai 201210, China}
\email{carol.guo@sjtu.edu.cn}

\begin{abstract}
As a byproduct of sub-Neptune formation, planetary embryos with high eccentricity can remain in outer orbits, near 1 au from the star. In this work, we investigate the long-term evolution of systems consisting of close-in sub-Neptunes (SNs) and outer high-eccentricity embryos. Our analysis focuses on collisions between SNs and embryos, particularly their atmospheric mass loss. We performed \textit{N}-body simulations for various initial eccentricities and numbers of embryos. We analyzed the impact-induced atmospheric loss using post-processing methods, finding that the embryos and SNs collide at high speeds on timescales of several million years, leading to the loss of the SNs’ atmospheres. Depending on the embryos’ eccentricity and the orbital radius of the SNs, the impact velocity can be quite high, ranging from 2 to 5 times the escape velocity. On average, $\sim$ 15\%--30\% of the atmosphere is dissipated per collision, so after $\sim$ 3--6 collisions, the atmospheric mass of an SN is reduced to about 1/3 of its initial value. Collisions between SNs and embryos can thus explain the presence of planets within the radius gap. Depending upon the initial eccentricity and the number of remaining embryos, additional collisions can occur, potentially accounting for the formation of the radius gap. This study also indicates that collisions between remaining embryos and SNs may help to explain the observed rarity of SNs with atmospheric mass fractions greater than 10\%, commonly termed the \lq\lq radius cliff.\rq\rq
\end{abstract}

\keywords{\uat{Super-Earths}{1655} --- \uat{Exoplanet formation}{492} --- \uat{Planetary atmospheres}{1244} --- \uat{N-body simulations}{1083}}


\section{Introduction}\label{sec:intro}
Super-Earths (SEs) and sub-Neptunes (SNs), collectively termed SENs which are with masses $\sim$ 2--10 $M_\oplus$ and sizes 2--3 $R_\oplus$, have been observed by the Kepler space telescope, and their orbital distributions, mass distributions, orbital spacings, eccentricities, and bulk compositions have been determined \citep[e.g.,][]{2023ASPC..534..839L, 2023ASPC..534..863W}. These data provide valuable insights into the origins of SENs.

Theoretical studies of the origin of SENs suggest that they typically grow at outer orbital locations and then undergo inward migration to close-in orbits after reaching masses $\sim 1 M_\oplus$ \citep{2017MNRAS.470.1750I,2022ApJ...939L..19I,2018A&A...612L...5O,2019A&A...627A..83L,2024ApJ...964L..13S,2024A&A...692A.246B}. It is debatable in which orbital region the SENs grow, but we can make an analogy. The occurrence of SENs is logarithmically uniform for orbital periods up to 300 days \citep{2018AJ....155...89P,2023AJ....166..122D}, suggesting that the formation region is situated around or outside the region with such orbital periods. Compositional estimates also suggest that many SENs likely do not contain large amounts of ice (i.e., $<10\%$ by mass) \citep{2017A&A...597A..37D,2019ApJ...883...79D,2020A&A...640A.135O,2021PSJ.....2....1A}. This makes it unlikely that most SENs formed beyond the snow line before migrating to their current close-in orbits \citep{2018MNRAS.479.4786V,2021MNRAS.503.1526R}. Recent studies have proposed planet formation from rings \citep[e.g.,][]{2016A&A...594A.105D,2021ApJ...914L..38U,2022ApJ...939L..19I,2022DPS....5410201M,2025ApJ...983...56G}, and models suggesting that SENs can form from rings at orbital distances of 1 au explain their observed properties \citep[e.g.,][]{2023NatAs...7..330B,2024ApJ...972..181O,2025ApJ...979L..23S}.

If SENs do grow and migrate inward, they settle in an inner region, while high-eccentricity embryos remain near the SEN-formation region \citep{2017A&A...607A..67M,2020ApJ...899...91O,2021A&A...650A.152I,2026ApJ...996...91O}. Those SENs with masses of a few $M_{\oplus}$ and embryos with about the mass of Mercury undergo close encounters during formation at $\sim$ 1 au from the star, resulting in an increase in the eccentricity of the smaller embryos. If the SENs form from a ring, high-eccentricity embryos may remain outside the ring. The SENs then undergo type I migration to inner orbits with semimajor axes $a \simeq 0.1\text{--}0.5\,\mathrm{au}$, while smaller high-eccentricity embryos remain in outer orbits around $a \simeq 1\,\mathrm{au}$, unaffected by orbital migration or eccentricity damping (see Section \ref{sec:sec4.2}). As a result, after the gas disk dissipates, high-eccentricity embryos may remain around $a \simeq 1$ au over the long term, undergoing subsequent collisions and scattering with the inner SENs. This can lead to changes in the orbital structure of the system and the properties of the SENs. However, the long-term evolution of such a system has not been rigorously studied previously \footnote{This is partly because high-resolution \textit{N}-body simulations starting from small embryos (0.01 $M_{\oplus}$) are required to capture the formation dynamics of high-eccentricity embryos properly \citep{2020ApJ...899...91O}. Such simulations are computationally expensive and difficult to follow from the growth phase to the long-term phase.}. The characteristics of collisions between SENs and embryos (e.g., the impact velocity and frequency) and whether they result in atmospheric loss from the SENs also remain largely unresolved.

The size distribution of SENs exhibits several remarkable features that provide critical constraints for planet-formation theories. One characteristic is the \lq\lq radius gap\rq\rq\ \citep[e.g.,][]{2017AJ....154..109F,2022AJ....163..179P,2023MNRAS.519.4056H}, which is a bimodal distribution, with peaks at $\sim$ 1.5 and 2.7 $R_\oplus$, separated by a valley at $\sim 2 R_\oplus$ \citep{2018MNRAS.479.4786V,2019ApJ...875...29M,2020AJ....160...89P}. Photoevaporation \citep[e.g.,][]{2012ApJ...761...59L,2013ApJ...775..105O,2017ApJ...847...29O,2018ApJ...853..163J} and core-powered mass loss \citep[e.g.,][]{2018MNRAS.476..759G,2019MNRAS.487...24G} have been widely studied as potential origins of the radius gap. Note that recent work shows that core-powered mass loss becomes much less efficient beyond the short boil-off phase, suggesting that its long-term role in sculpting the radius valley may be limited \citep{2024ApJ...976..221T}. Moreover, neither process explains the log-uniform period distribution of SENs easily \citep{2018AJ....155...89P,2023AJ....166..122D}. Furthermore, if future observations show that a radius gap exists for planets at long orbital periods where stellar irradiation is weak, it would challenge the irradiation-driven model of photoevaporation. In addition, the photoevaporation model cannot easily explain the observed size non-uniformity in systems that have planets within the gap \citep{2024arXiv241002150C}. It has also been suggested that planets within the gap may also have higher eccentricities \citep{2025PNAS..12205295G}, and the photoevaporation model cannot account for this. Other potential mechanisms to explain the radius gap include composition differences \citep{2019PNAS..116.9723Z,2020A&A...643L...1V,2022ApJ...939L..19I} and impact erosion \citep{2015MNRAS.448.1751I,2020ApJ...892..124O,2021ApJ...923...81M,2022ApJ...937...39C}. Which mechanisms work and to what extent thus remains an important topic in radius-gap studies.

Another notable feature is that few SENs have radii $> 3.5 R_\oplus$ (corresponding to planets with hydrogen-dominated atmospheres exceeding 10\% of their total mass atop rocky cores), especially compared to the occurrence rate of SENs with radii $< 3.5 R_\oplus$ \citep[e.g.,][]{2019AJ....158..109H}. However, it is theoretically possible for SENs to acquire atmospheres $\sim$ 10\% of the planetary mass, similar to those of Uranus and Neptune \citep[e.g.,][]{2014ApJ...786...21P}. Several mechanisms have been proposed to avoid the formation of SENs with atmospheres exceeding 10\% of the mass, including high atmospheric opacity \citep{2014ApJ...797...95L,2020A&A...634A..15B}, rapid recycling \citep{2015MNRAS.447.3512O,2017A&A...606A.146L,2019A&A...623A.179K}, limitations imposed by disk accretion \citep{2018ApJ...867..127O,2019MNRAS.490.4334G,2020ApJ...899...91O}, atmospheric loss during disk dissipation or boil-off \citep{2012ApJ...753...66I,2016ApJ...817..107O,2020ApJ...889...77H,2024ApJ...976..221T}, and the fugacity crisis \citep{2019ApJ...887L..33K}, but this topic remains under debate. The so-called \lq\lq radius cliff\rq\rq\ is also difficult to explain by photoevaporation or core-powered mass-loss models that are tuned to explain the radius valley, as these models underpredict the abundance of large SNs (>3 $R_\oplus$) at long orbital periods, and they exhibit inconsistent slope variations across stellar types \citep{2024AJ....167..288D}.

Collisions between SENs and embryos that occur during the long-term evolution of a system with inner SENs and outer high-eccentricity embryos---which we consider here---alter the size distribution of the SENs. This mechanism may explain the origins of planets inside and outside the radius gap. Although some investigations have studied the effect on the radius gap of giant impacts between SENs \citep{2021ApJ...923...81M,2022ApJ...937...39C}, the role of impact erosion between SENs and embryos in forming planets within the radius gap and in shaping the radius gap is currently unknown. 

In the present study, we use \textit{N}-body simulations to investigate the long-term evolution of a system of inner SNs and outer high-eccentricity embryos after disk dissipation. We consider SENs with atmospheres as the initial conditions for the simulations; we refer to them hereafter as SNs and those without atmospheres as SEs. Our goals are to clarify how the system evolves, study the properties of collisions between SNs and embryos (e.g., frequency, velocity, and impact angle), and determine the amount of atmospheric mass loss due to the collisions. We also investigate whether planets can form in the radius gap and how such a process helps shape the radius gap. Our results also have implications for the origin of the radius cliff. Because it is computationally expensive to follow the growth and long-term evolution of SENs using \textit{N}-body simulations, in this study, we consider the long-term evolution only from the stage when their growth and migration are complete and the gas disk has dissipated. We also ignore the gravitational interactions among the embryos in most of the simulations in order to speed up the simulations.

The paper is organized as follows: Section \ref{sec:model} describes the initial conditions and numerical methods. Section \ref{sec:results} presents the simulation results and analyses. Section \ref{sec:Discussion} discusses the validity of our test-particle simulations, our initial conditions, and the caveats of the model. Finally, Section \ref{sec:Conclusions} summarizes our conclusions.

\section{Model} \label{sec:model}
\subsection{Initial conditions}
Our model considers three SNs in the inner region \citep{2018ApJ...860..101Z} and many high-eccentricity planetary embryos in the outer region. To study the long-term evolution of the system, we perform a series of simulations with different initial conditions, varying the number and eccentricities of the planetary embryos. The semimajor axis of the innermost SN is taken from the log-normal distribution,  
\begin{equation}
f(x) = \frac{1}{x\sigma\sqrt{2\pi}} \exp\left(-\frac{(\ln x - \mu)^{2}}{2\sigma^{2}}\right), 
\end{equation}
with a mean value of 0.15 au [$\mu=\ln(0.15\, \text{au})$] and a standard deviation of 0.25 au [$\sigma=0.25$], where $x$ represents the semimajor axis \citep{2017AJ....154..109F,2022AJ....163..179P}. The second and third SNs are placed with orbital separations of 30 times the mutual Hill radius
\citep{2014ApJ...784...44L,2023ASPC..534..863W}. This setup reproduces an approximately log-uniform occurrence distribution for orbital periods beyond about 20 days. The masses of the three SNs ($M_\text{core}$) are assumed to follow the log-normal distribution adopted in Model I of \citet{2021MNRAS.503.1526R}, with a mean of 3$M_\oplus$ and a standard deviation of 0.4. We note, however, that the true core-mass distribution is still poorly constrained. More flexible models in \citet{2021MNRAS.503.1526R} yield a broader distribution, and such inferences are sensitive to forward-evolution assumptions that omit several important physical processes such as early boil-off \citep[e.g.,][]{2025arXiv251002201T}, volatile dissolution \citep{2018ApJ...854...21C,2019ApJ...887L..33K,2025MNRAS.tmp.1830R}, core thermal evolution \citep{2018ApJ...869..163V,2025ApJ...989...28T}, and atmospheric metallicity effects \citep{2010ApJ...712..974R,2019ApJ...874L..31T,2025arXiv251002201T,2025ApJ...989...28T}. Our adopted distribution should therefore be regarded as a reasonable but simplified choice for population-level modeling. The eccentricity and inclination of the SNs are set to 0.02 and 0.01 rad, respectively \citep{2016PNAS..11311431X,2019AJ....157..198M,2019AJ....157...61V}. The longitude of the ascending node, the argument of periapsis, and the true anomaly are random numbers ranging from 0 to 2$\pi$. The three SNs remain in stable orbits throughout the evolution of the system.

Planetary embryos with masses $M_\text{emb}= 0.05 M_\oplus$ are distributed uniformly in a two-dimensional plane with their semimajor axes between 1 au and 2 au. The orbital longitude of the ascending node, the argument of periapsis, and the true anomaly are random numbers ranging from 0 to 2$\pi$. The eccentricities ($e_\text{ini}$) are set to 0.7, 0.8, and 0.9, and the numbers of embryos ($N_\text{ini}$) are set to 60, 80, and 100 \citep{2020ApJ...899...91O,2022ApJ...939L..19I}. By combining different eccentricities with different numbers of embryos, we get a total of nine different models, and we performed 20 simulations for each model case to ensure statistical robustness.

\subsection{\textit{N}-body simulations}\label{sec:n-body}
We ran our \textit{N}-body simulations with the REBOUND code \citep{2012A&A...537A.128R}, employing the MERCURIUS integrator \citep{2019MNRAS.489.4632R} with the time step set to 1/40 of the orbital period at a semimajor axis of 0.1 au. We ran each simulation for a total duration of 50 million years. To verify that our adopted integration time is sufficient, we extended the fiducial set of simulations ($N_{\text{ini}} = 80 $, $e_{\text{ini}} = 0.8$) to 100 million years and found that the results change only marginally. We treat the embryos as test particles, considering only the gravitational interaction between the embryos and the SNs and ignoring the mutual interactions among the embryos. To check the validity of our results, we have also performed full \textit{N}-body simulations that account for the gravitational interactions among all celestial bodies (Section \ref{sec:sec4.1}).

The planetary radius is calculated from the planet’s core mass and density ($4.5\,\mathrm{g/cm^3}$) \citep{2021ApJ...908...32L}. If the distance between the centers of two celestial bodies is less than the sum of their radii, they are considered to merge. This treatment ignores grazing impacts that interact only with the atmosphere without reaching the solid core, which is consistent with the impact‐erosion model adopted in this study (described below). Although the influence of atmospheric and core grazing impacts is thought to be limited \citep{2022MNRAS.513.1680D}, incorporating such collisions will be an important topic for future work. After a collision, the mass of the SN is increased by the mass of the accreted embryo, and this updated mass is used in all subsequent \textit{N}-body interactions. To determine whether a planet has been scattered out of the system, we calculate whether the distance of the planet from the center of mass of the system is $>10^6$ au. We have confirmed that when the distance exceeds this threshold, it no longer affects the system, and we remove it.

\subsection{Post-processing analysis of atmospheric loss}\label{sec:post}
The simulations record position and velocity information during collisions between planetary embryos and SNs, enabling calculations of the impact velocity and angle. With this information, we can calculate the fraction of the atmospheric mass loss during each collision. Subsequently, we calculate the final radius of the planets using a formula provided by \citet{2014ApJ...792....1L}. 
Unless otherwise specified, we set the initial atmospheric mass fraction of each SN to follow a log-normal distribution with a mean of 3\% and a standard deviation of 0.3, which is analogous to that in previous studies \citep[e.g.,][]{2021MNRAS.503.1526R}.

To obtain the atmospheric mass fraction $f_\text{atm}$ for each SN, we calculate the fractional atmospheric mass loss $X$ caused by all the collisions \citep{2020ApJ...901L..31K,2021ApJ...923...81M} . We used the scaling law derived by \citet{2020ApJ...901L..31K}, which models atmospheric loss driven by ground motion during impacts, based on SPH impact simulations. In addition to such kinetic effects, thermal effects may also play some role in atmospheric escape \citep[e.g.,][]{2019MNRAS.485.4454B}.\footnote{Thermal effects may include processes such as mantle melting, magma-ocean formation, core–mantle differentiation, and atmospheric degassing \citep[e.g.,][]{1986Natur.319..303M,1993JGR....98.5319T,2016PEPS....3....7D,2023Icar..40615739D}. Previous studies have primarily examined collisions where the target and impactor do not differ greatly in mass. In contrast, the regime considered in this study involves somewhat smaller impactors, and how the associated thermal evolution proceeds in such cases remains an interesting subject for future investigation.} However, because an appropriate 3D thermal–kinetic model is not available for the collision regime examined here, we limit our post-processing analysis to the kinetic component alone and adopt the scaling relation. Following \citet{2020ApJ...901L..31K}, we use
\begin{equation}
X \approx 0.64 \left[ \left( \frac{v_{\text{imp}}}{v_{\text{esc}}} \right)^2 \left( \frac{M_\text{emb}}{M_{\text{tot}}} \right)^{\frac{1}{2}} \left( \frac{\rho_\text{emb}}{\rho_\text{core}} \right)^{\frac{1}{2}} f_M(b) \right]^{0.65},
\label{eq:eq1}
\end{equation}
where the impact velocity $v_{\text{imp}}$ represents the relative velocity between embryo and SN at the time of collision. The escape velocity is $v_{\text{esc}} = (2GM_{\text{tot}}/R_{\text{tot}})^{1/2}$, where $M_{\text{tot}} = M_\text{emb} + M_\text{core}$ and $R_{\text{tot}} = R_\text{emb} + R_\text{core}$. The radius and density of the embryo are $R_\text{emb}$ and $\rho_\text{emb}$, respectively, and the corresponding properties of the SN are $R_\text{core}$ and $\rho_\text{core}$. The mass $M_\text{core}$ used here corresponds to the mass of the SN at the instant of first contact, i.e., immediately before the merge occurs. After the merge, the increased $M_\text{core}$ is used in the calculation of the next collision. The mass fraction $f_M$ is defined as the ratio of the total mass involved in the collision to the combined total mass of both bodies (see the shaded region in Fig.~7 of  \citealt{2012ApJ...744..137G}). Mathematically, it is given by
\begin{equation}
f_M=\frac{\rho_\text{core} V_\text{core}^{\text{cap}} + \rho_\text{emb} V_\text{emb}^{\text{cap}}} {\rho_\text{core} V_\text{core} + \rho_\text{emb} V_\text{emb}},
\end{equation}
where $V_\text{core}$ and $V_\text{emb}$ are the total volumes of the SN and the embryo, respectively, and the numerator represents the mass contained in the interacting caps.
If $\rho_\text{core}=\rho_\text{emb}$, then 
\begin{eqnarray}
f_M &=&\frac{V_\text{core}^{\text{cap}} + V_\text{emb}^{\text{cap}}}{V_\text{core} + V_\text{emb}} \nonumber\\
&=& 0.25 \frac{(R_\text{core} + R_\text{emb})^3}{R_\text{core}^3 + R_\text{emb}^3} (1 - b)^2 (1 + 2b), 
\end{eqnarray}
where $b = \sin \theta$ is the dimensionless impact parameter. The volume $V_\text{core}^{\text{cap}}$ represents the portion of the SN cap above the lowest point of the embryo at contact and $V_\text{emb}^{\text{cap}}$ the portion of the embryo cap below the highest point of the SN (see the schematic illustration in Fig.~1 of \citealt{2020ApJ...897..161K}). Both caps have height $d = (R_\text{core} + R_\text{emb})(1 - b)$, giving 
\begin{equation}
V_\text{core,emb}^{\text{cap}} = \frac{\pi}{3} d^2 (3R_\text{core,emb} - d).
\end{equation}
As Equation (\ref{eq:eq1}) shows, the factor that influences the proportion of atmospheric loss most is the impact velocity between the planetary embryo and the SN.

Once we have analyzed all the collision events, we compute the final radius of each SN. Following \citet{2014ApJ...792....1L}, the planetary radius can be calculated as:
\begin{eqnarray}
R_\text{SN} &=& R_{\text{env}} + R_{\text{core}} + R_{\text{atm}}, \label{eq:R_SN}\\
R_{\text{env}} &=& 1.06 R_{\oplus}\left(\frac{M_{\rm core}}{3 M_{\oplus}}\right)^{-0.21}
                  \left(\frac{f_{\text{atm}}}{3\%}\right)^{0.59} \nonumber \\
              && \times\left(\frac{F_\text{SN}}{F_{\oplus}}\right)^{0.044}
                  \left(\frac{\text{age}}{10\,\text{Gyr}}\right)^{-0.18}, \\
R_{\text{core}} &=& 1.32 R_{\oplus} \left(\frac{M_{\rm core}}{3 M_{\oplus}}\right)^{0.25},\\
R_{\text{atm}} &\simeq& 9\left(\frac{k_{b}T_{eq}}{g\mu_{{\rm H/He}}}\right).\label{eq:radius}
\end{eqnarray}
The planet consists of three radial components: the solid core with radius $R_{\text{core}}$, the convective H/He envelope $R_{\text{env}}$, and the radiative atmosphere $R_{\text{atm}}$ (typically defined at a pressure of 20 mbar). The term atmosphere here refers specifically to the radiative layer ($R_{\text{atm}}$). In contrast, elsewhere in this paper, particularly when discussing atmospheric mass fractions $f_{\rm atm}$ and atmospheric loss, the term atmosphere refers to the convective envelope.
The symbols $F_\text{SN}$, $k_b$, $T_{eq}$, $g$, and $\mu_{\text{H/He}}$ represent the incident flux, Boltzmann’s constant, the equilibrium temperature, the gravity of the SN, and the mean molecular weight, respectively. 
This set of Equations~(\ref{eq:R_SN})-(\ref{eq:radius}) is intended to reproduce the one dimensional structure calculations of \citet{2014ApJ...792....1L}. However, we find that if we adopt $T_{\rm eq} \propto r^{-1/2}$ and take $g$ at the top of the envelope ($R_{\rm env}$), the atmosphere layer $R_{\rm atm}$ becomes too large and the resulting radii are inconsistent with the structure calculations of \citet{2014ApJ...792....1L}. Therefore, in this study we compute the planet size as $R_{\rm SN} = R_{\rm core} + R_{\rm env}$, which is more consistent with the results of \citet{2014ApJ...792....1L}. Note that recent structure calculations suggest that the radiative atmosphere may in fact be thicker (see Section~\ref{sec:radius_model_limits} for discussion).
We define the orbital distance of the SN as $r$ and assume $F_\text{SN}/F_{\oplus} = \left[ r/(1\,\text{au}) \right]^{-2}$ and $\text{age} = 10\,\text{Gyr}$.

\section{Results} \label{sec:results}
\subsection{Collision Dynamics and Atmospheric Mass Loss}
\subsubsection{Dynamical Evolution in a Typical Simulation}
\begin{figure*}[ht!]
\plotone{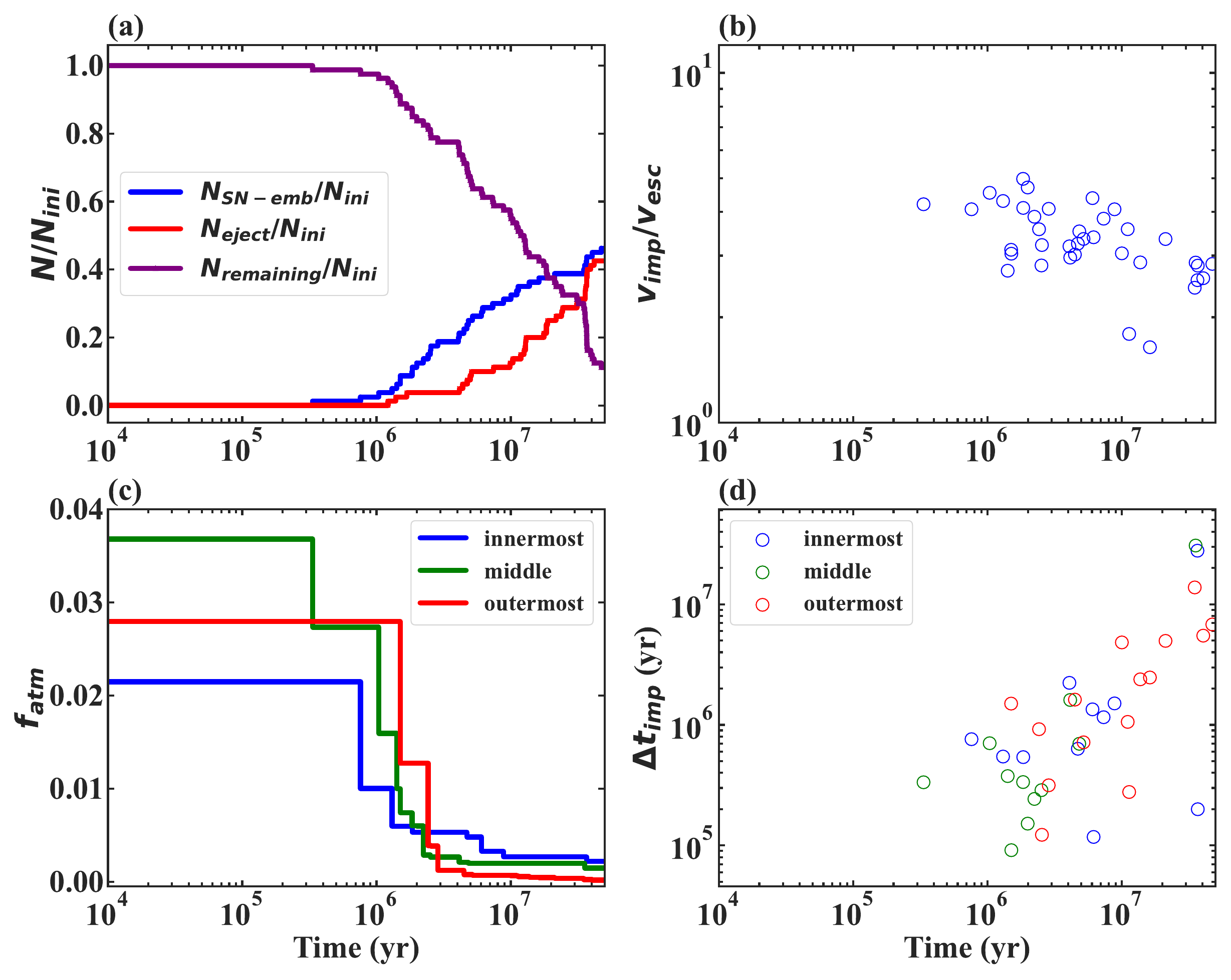}
\caption{Time evolution of our typical simulation for $N_{\text{ini}} = 80 $ with $e_{\text{ini}} = 0.8$. (a) The blue line represents the collisions between SNs and embryos, the red line represents the ejection of embryos, and the purple line represents the remaining embryos. (b) The impact velocities between SNs and embryos. (c) The atmospheric mass fractions of the innermost, middle, and outermost SNs are represented by the blue, green, and red lines, respectively. (d) Time intervals between successive SN–embryo collisions for the three SNs (blue, green, and red circles), showing that the impact intervals increase over time, indicating longer impact-erosion timescales at later stages.
\label{fig:fig1}}
\end{figure*}

We first discuss a simulation with $N_{\text{ini}} = 80$ embryos and eccentricity $e_\text{ini} = 0.8$. Figure \ref{fig:fig1} shows a typical result of one of the 20 simulation runs. We show simulations with different eccentricities and different numbers of embryos later, but the qualitative evolution of the system is the same in all cases.

Figure \ref{fig:fig1}(a) shows the time evolution of the cumulative number $N_{\text{SN-emb}}$ of collisions of embryos with SNs, the number $N_{\text{eject}}$ of embryos ejected from the system, and the number $N_{\text{remaining}}$ of remaining bodies. Most collisions between SNs and embryos occur after 1 Myr. By the end of the simulation, 48\% of the embryos have collided with SNs, and 41\% have been ejected from the system.\footnote{To confirm that an integration time of 50 Myr is sufficient, we extended all 20 runs of this fiducial case to 100 Myr. The total number of collisions between SNs and embryos then increases only from 38 to 39 (averaged over 20 runs), showing that late-time impacts are rare and have a negligible effect on the atmospheric mass.} There are no collisions between SNs. Figure \ref{fig:fig1}(b) shows the impact velocities during the collisions between SNs and embryos. The impact velocity can be quite high, mostly ranging from 2 to 5 times the escape velocity. Figure \ref{fig:fig1}(c) shows the atmospheric mass fractions of the three SNs. We find that large amounts of the atmosphere are lost due to SN--embryo collisions; the atmospheric mass fractions change from $\simeq 3\%$ to $\simeq 0.2\%$ in 10 $\mathrm{Myr}$. In this figure, we also find a difference from the photoevaporation model. Specifically, all three SNs experience impact-driven atmospheric erosion. For the outermost planet ($a_\mathrm{SN} \simeq 0.45\ \text{au}$), the contrast with photoevaporation is clear, because photoevaporative mass loss is expected to be weak at such orbital distances under our adopted assumptions.

We further identify differences from the photoevaporation model.
In impact erosion, the amount of mass loss is proportional to $X f_{\rm atm}$; that is, the greater the amount of atmosphere $f_{\rm atm}$ retained at the time of impact, the larger is the absolute amount of atmospheric loss. Therefore, as time passes and the atmosphere becomes thinner, atmospheric mass loss due to impact erosion decreases. 
The difference becomes clear when we compare the characteristic timescales. In photoevaporation, the mass-loss timescale is long when the atmospheric mass fraction is a few percent, but it becomes rapidly shorter once the atmosphere becomes very thin, for example around 0.01 to 0.1 percent \citep[e.g.,][]{2019AREPS..47...67O}.
This behavior is qualitatively different from atmospheric loss caused by impact erosion. In contrast, the characteristic timescale for impact erosion is given by
\begin{equation}
    \tau \equiv - \frac{f_{\rm atm}}{df_{\rm atm}/dt} = \frac{\Delta t_{\rm imp}}{|\ln(1-X)|} \simeq \frac{\Delta t_{\rm imp}}{X},
\end{equation}
where $\Delta t_{\rm imp}$ is the impact interval. Since the fractional loss per impact $X$ does not depend strongly on time, the timescale $\tau$ is mainly determined by the impact interval. Figure~\ref{fig:fig1}(d) shows the impact intervals for each SN, and these intervals increase with time. This indicates that the timescale for impact erosion increases as the system evolves.

\subsubsection{Dependence of Mass Loss from a Single Collision on Orbital Distance and Eccentricity}
\begin{figure*}[ht!]
\plotone{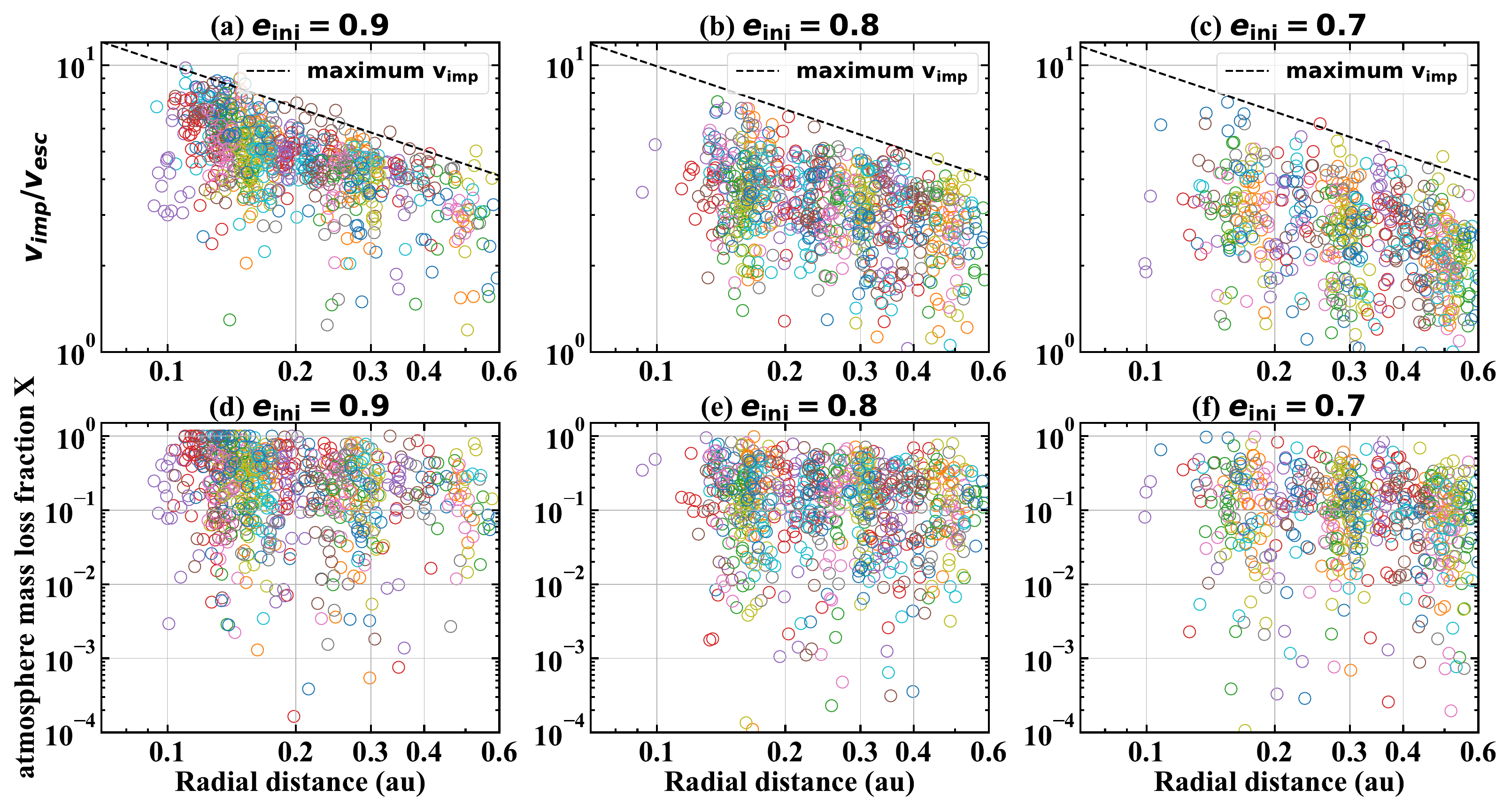}
\caption{Summary of collisions between SNs and embryos in all simulations with $N_{\text{ini}}$=80 and different initial embryo eccentricities. (a)--(c) Impact velocities when SNs collide with embryos. The dashed line is a rough estimate of the maximum impact velocity. (d)--(f) Fractional atmospheric mass loss during collisions between SNs and embryos. The same colors represent collisions within the same system.  
\label{fig:fig2}}
\end{figure*}

We next analyze the collision velocities and atmospheric mass-loss fraction of all collisions between SNs and embryos in the 80-embryo case, with initial embryo eccentricities $e_\text{ini} = 0.9, 0.8,$ and 0.7. Figure \ref{fig:fig2} compiles results from 20 simulations per setting. If an embryo with semimajor axis $a$ collides at position $r$ with an SN on a circular orbit, the impact velocity $v_{\rm imp}$ is approximately given by
\begin{eqnarray}
v_\text{imp}^2 &=& v_\text{emb}^2 + v_K^2 \nonumber\\
&=& G(M_{\odot} + M_\text{emb}) \left(\frac{2}{r} -\frac{1}{a} \right) + \left( \frac{GM_{\odot}}{r} \right).
\label{v_imp}
\end{eqnarray}
We assume that the collision happens at the pericenter of the embryo, $a = r/(1-e)$, and use the escape velocity of the SN, $v_{\rm esc, core} = \sqrt{2GM_{\rm core}/R_{\rm core}}$. Then
\begin{equation}\label{eq:vimp}
    \frac{v_\text{imp}}{v_\text{esc, core}} \simeq \sqrt{\frac{(2+e) M_\odot R_{\rm core}}{2 M_{\rm core} r}}.
\end{equation}
As Figure \ref{fig:fig1} shows, SN--embryo collisions typically occur at 2--5 times the escape velocity in these three cases [Figures~\ref{fig:fig2}(a)--(c)], dissipating $\sim$ 15\%--30\% in a single collision [Figures~\ref{fig:fig2}(d)--(f)].

Some collisions can reach the maximum impact velocity given by Equation~(\ref{eq:vimp}). Collisions near the system center achieve the highest velocities because the orbital velocities of both the SNs ($v_K = \sqrt{GM_{\odot}/r}$) and embryos [$v_\text{emb} = \sqrt{G(M_{\odot}+M_\text{emb})(2/r - 1/a)}$] peak at small $r$. This explains why inner systems show greater atmospheric stripping. We also find that some SNs with $a_\mathrm{SN} > 0.4$ au (i.e., orbital periods $P > 100$ days) can undergo nearly complete atmospheric loss ($X_{\rm tot}\gtrsim 0.9$), which is difficult to achieve via photoevaporation alone under our adopted conditions (i.e., the initial envelope mass fraction and the SN size).

Figures \ref{fig:fig2}(a)--(c) and Figures \ref{fig:fig2}(d)--(f) show the dependence on the embryos' initial eccentricity. Higher eccentricities allow embryos to collide with SNs at closer distances; thus, $e_\text{ini} = 0.9$ enables many collisions to occur in the inner region ($r \simeq 0.1\,{\rm au}$). Consequently, the arithmetic mean values of the impact velocity and atmospheric mass-loss fraction are higher than they are for $e_\text{ini} = 0.7$. In Table~\ref{tab:tab1}, the average impact velocity is 4.93 $v_{\rm esc}$ for $e_{\rm ini} = 0.9$, while it is 2.73 $v_{\rm esc}$  for $e_{\rm ini} = 0.7$.

Because the mass-loss fraction $X$ depends on the impact velocity [Equation (\ref{eq:eq1})], the average mass loss is higher in simulations with higher initial eccentricities and for SNs at smaller orbital distances. 
The average atmospheric mass-loss fraction $\overline{X}$ increases with initial eccentricity, as shown in Table \ref{tab:tab1}.

\subsubsection{Dependence of Total Atmospheric Loss on Collision Number and Eccentricity}
\begin{deluxetable*}{lccccccc}
\tabletypesize{\footnotesize}
\tablecaption{Averaged value of collision and atmospheric mass loss for each setting\label{tab:tab1}}
\tablewidth{0pt} % Natural width
\tablehead{
\colhead{$e_{\mathrm{ini}}$} & 
\colhead{$N_{\mathrm{ini}}$} & 
\colhead{$\overline{{v}_\mathrm{imp}}$} & 
\colhead{$\overline{{N}_\mathrm{SN-emb}}$} & 
\colhead{$\overline{{N}_\mathrm{eject}}$} & 
\colhead{$\overline{X}$} & 
\colhead{$\overline{{X}_\mathrm{tot}}$} & 
\colhead{$F_{X_{\mathrm{tot}}<0.9}$} \\
\colhead{} & 
\colhead{} & 
\colhead{($v_{\mathrm{esc}}$)} & 
\colhead{} & 
\colhead{} & 
\colhead{(\%)} & 
\colhead{(\%)} & 
\colhead{(\%)}
}
\startdata
0.9 & 100 & 4.85 & 17 & 44 & 29.0 & 89.5 & 21.7\\
0.9 & 80 & 4.93 & 13 & 36 & 29.9 & 88.9 & 30.0 \\
0.9 & 60 & 4.89 & 10 & 26 & 29.2 & 82.9 & 38.3\\
0.8 & 100 & 3.32 & 16 & 41 & 19.1 & 91.1 & 23.3\\
0.8 & 80 & 3.38 & 13 & 33 & 20.3 & 88.6 & 31.7 \\
0.8 & 60 & 3.38 & 10 & 25 & 20.0 & 82.6 & 46.7 \\
0.7 & 100 & 2.64 & 12 & 27 & 15.7 & 83.3 & 43.3\\
0.7 & 80 & 2.73 & 10 & 21 & 15.3 & 76.7 & 73.3 \\
0.7 & 60 & 2.64 & 7 & 16 & 14.8 & 67.9 & 88.3\\
0.8 & 80 & 3.71 & 10 & 25 & 23.8 & 87.4 & 38.3 \\
\enddata
\tablecomments{For simulations of systems with varying initial embryo eccentricities ($e_\mathrm{ini}$) and initial embryo numbers ($N_{\mathrm{ini}}$), this table lists the mean impact velocity ($\overline{{v}_\mathrm{imp}}$), average number of collisions between SNs and embryos ($\overline{{N}_\mathrm{SN-emb}}$), the number of ejected embryos ($\overline{{N}_\mathrm{eject}}$), the average percentage of atmospheric mass lost per collision ($\overline{X}$), the cumulative percentage of atmospheric mass lost per SN ($\overline{{X}_\mathrm{tot}}$), and the fraction of SNs with total atmospheric mass loss less than 90\% ($F_{X_{\mathrm{tot}}<0.9}$). The final row shows results from a full \textit{N}-body simulation for a typical case ($e_\mathrm{ini}=0.8$, $N_{\mathrm{ini}}=80$), enabling comparison with the test simulation.}
\end{deluxetable*}

Table \ref{tab:tab1} summarizes the results of nine simulation sets with initial embryo eccentricities $e_\text{ini} = 0.9, 0.8,$ and $0.7$, and initial embryo numbers $N_\text{ini} = 60, 80,$ and $100$. For cases with the same initial eccentricity, the impact velocities are similar. As noted above, a higher initial eccentricity leads to a larger average impact velocity $\overline{v_\mathrm{imp}}$, which in turn increases $\overline X$.
During the 50-Myr simulations, multiple collisions occur between SNs and embryos. For a fixed initial number of embryos $N_{\mathrm{ini}}$, the average number of SN--embryo collisions $\overline{{N}_\mathrm{SN-emb}}$ decreases systematically with decreasing initial embryo eccentricity $e_{\mathrm{ini}}$ (Table~\ref{tab:tab1}). For example, for $N_{\mathrm{ini}}=80$, $\overline{{N}_\mathrm{SN-emb}}$ decreases from 13 for $e_{\mathrm{ini}}=0.9$ to 10 for $e_{\mathrm{ini}}=0.7$. This reflects the reduced orbital crossing probability and encounter rate at lower eccentricities. A larger initial number of embryos ($N_\text{ini}$) results in a higher average number of collisions between SNs and embryos ($\overline{N_\mathrm{SN-emb}}$). While the average atmospheric mass loss per collision ($\overline{X}$) is comparable across the simulations for a fixed $e_{\mathrm{ini}}$, a higher collision frequency yields a larger cumulative atmospheric mass loss per SN ($\overline{X_\text{tot}}$). In addition, systems with higher initial eccentricities experience both more SN--embryo collisions and higher impact velocities, leading to a systematically larger total atmospheric mass loss. Consequently, the fraction of SNs retaining more than 10\% of their atmospheres ($F_{X_{\text{tot}} < 0.9}$) decreases, indicating that more SNs are left with only thin atmospheres.

\subsubsection{Collision Frequency and Remaining Atmosphere as Functions of Radial Distance and Eccentricity}
\begin{figure*}[ht!]
\plotone{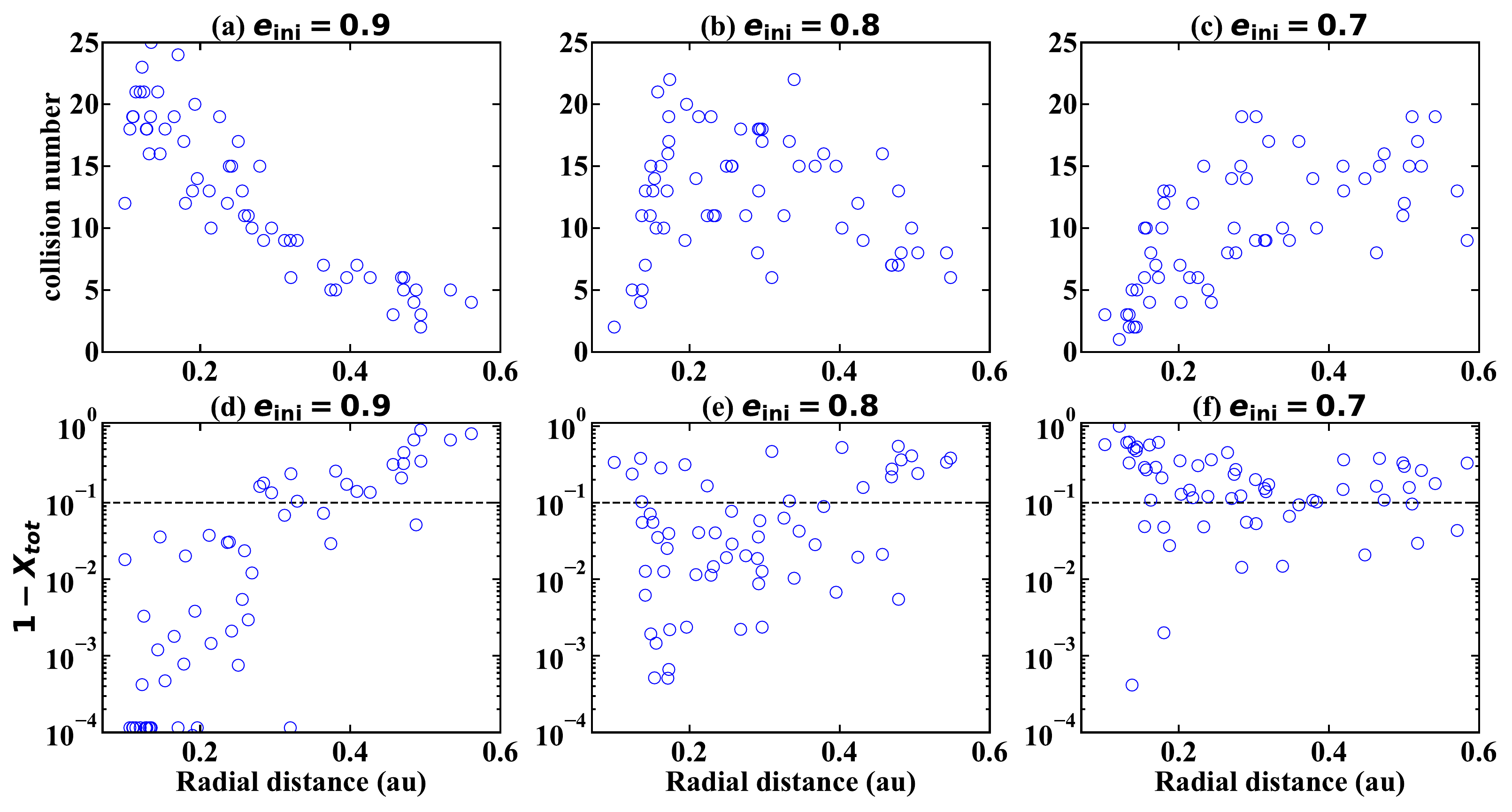}
\caption{Relationship between the positions of SNs and (a)--(c) the number of collisions experienced, and (d)--(f) the remaining atmospheric mass fraction. In panel (d), a value of $10^{-4}$ corresponds to complete atmospheric stripping. The dashed horizontal line marks a remaining atmospheric mass equal to 10\% of the initial atmospheric mass.
\label{fig:fig3}}
\end{figure*}

Figure~\ref{fig:fig3} presents the collision statistics and final atmospheric mass fractions, and their dependence on radial distance, for systems with 80 embryos and the eccentricities $e_\text{ini}=0.7$, 0.8, and 0.9. 

The orbital locations with high collision frequencies appear to depend on the initial eccentricity of the embryos.
For $e_\text{ini}=0.9$, the collision rate reaches its maximum at orbital semimajor axes $a_\mathrm{SN} \simeq 0.1$--0.2 au [Figure~\ref{fig:fig3}(a)], which coincides with 
the pericenter distance of embryos initially located near 1 au, $1~\mathrm{au} \times (1 - 0.9) = 0.1~\mathrm{au}$. This peak shifts to $a_\mathrm{SN} \simeq 0.2$--0.3 au when $e_\text{ini}=0.8$ [Figure~\ref{fig:fig3}(b)], consistent with the embryo pericenter $1~\mathrm{au} \times (1 - 0.8) = 0.2~\mathrm{au}$. For $e_\text{ini}=0.7$, the peak moves further outward to $a_\mathrm{SN} \simeq 0.3$--0.5 au [Figure~\ref{fig:fig3}(c)], matching the embryo pericenter $1~\mathrm{au} \times (1 - 0.7) = 0.3~\mathrm{au}$. The decline in collision frequency at small orbital distances occurs because the orbit of the SN lies inside the pericenter of the innermost embryos, so that orbital crossing is suppressed. Collisions are therefore concentrated near the embryo pericenters, although the full $N$-body simulations (Figure~\ref{fig:fig6}) exhibit a weaker radial dependence because the eccentricities of the embryos evolve.

The slope of the remaining atmospheric fraction $(1-X_\text{tot})$ as a function of orbital distance also depends on $e_{\text{ini}}$. 
With larger $e_{\text{ini}}$, $(1-X_\text{tot})$ decreases more steeply toward smaller orbital distances because collisions are more frequent and the impact velocity is higher [Figure~\ref{fig:fig3}(d)].
On the other hand, for smaller $e_{\text{ini}}$ [Figure \ref{fig:fig3}(f)], the slope of $(1-X_\text{tot})$ as a function of orbital distance becomes much shallower because of two competing effects: collisions are less frequent but the impact velocity is higher in the close-in region. Thus, the remaining atmospheric fraction exhibits a strong radial dependence only when high-eccentricity embryos are present in the system. However, the gravitational interaction between the embryos can weaken this dependence (Figure \ref{fig:fig6}).

For $e_\text{ini} = 0.9$ [Figure~\ref{fig:fig3}(d)], almost all planets except the outer planets ($a_\mathrm{SN} > 0.3$ au) lose most of their atmosphere ($X_{\text{tot}} > 0.9$). Only about 30\% of the SNs retain more than 10\% of their initial atmospheric mass (Table \ref{tab:tab1}). As the initial atmospheric mass fraction is about 3\%, these SNs retain atmospheric masses $\ge 0.3\%$. For $e_\text{ini} = 0.8$ [Figure~\ref{fig:fig3}(e)], about 69\% of the SNs lose 90\% or more of their initial atmosphere. The fraction of planets that retain more than 10\% of their initial atmosphere is still small (31\%) (Table \ref{tab:tab1}), and the cumulative average mass-loss fraction is $\overline{X_\text{tot}}= 88.6\%$. For $e_\text{ini} = 0.7 $ [Figure~\ref{fig:fig3}(f)], only about 30\% of the SNs lose 90\% or more of their atmospheres. In summary, for $N_\text{ini} = 80$, most of the atmospheres of SNs are dissipated for cases with $e_\text{ini} = 0.8$ and 0.9.

\subsection{Radius and Orbital-Period Distributions}
\begin{figure*}[ht!]
\plotone{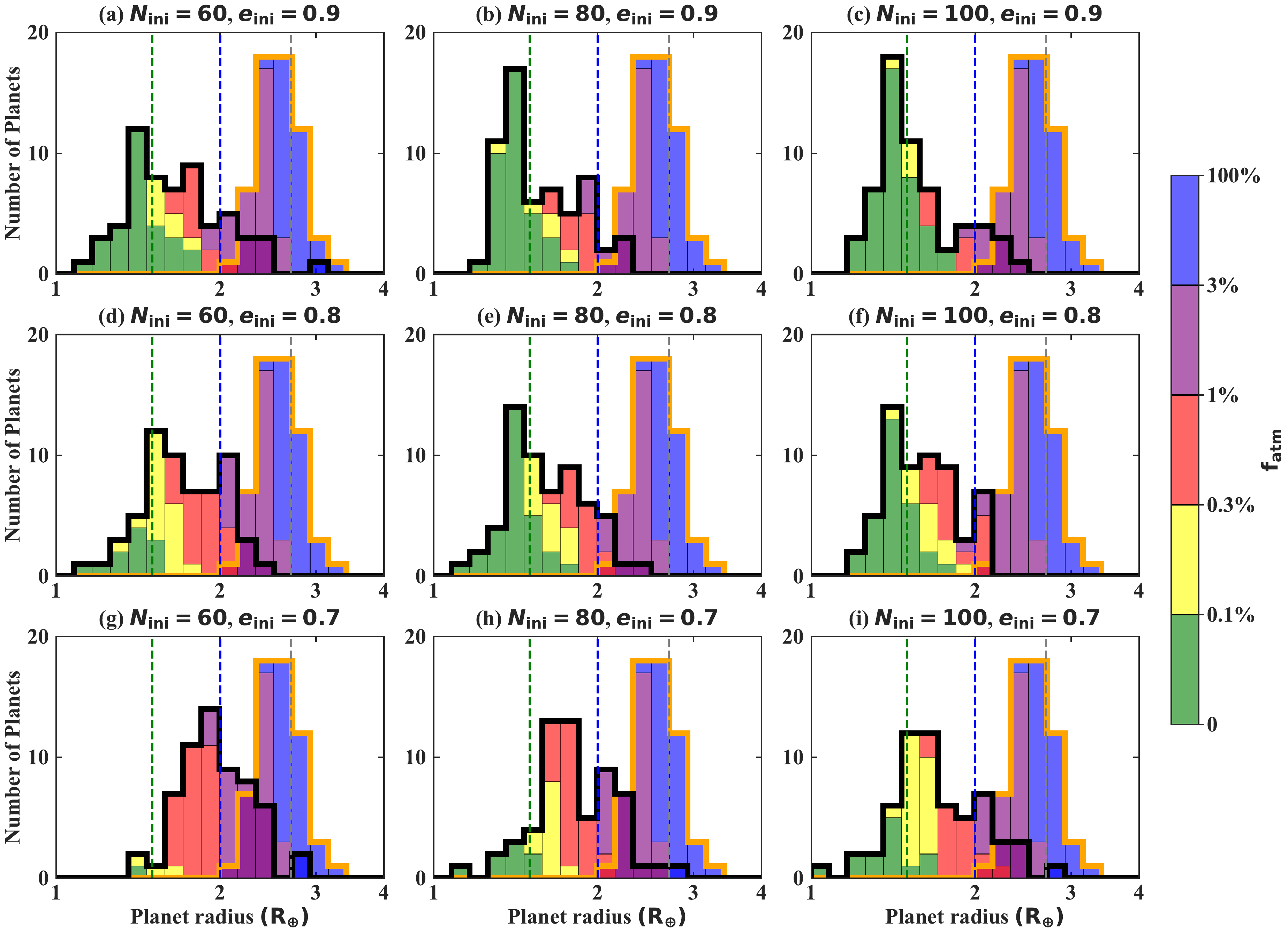}
\caption{The initial radius distribution (thick orange histogram) and final radius distribution (thick black histogram) of SNs after 50 Myr across different models. Histograms of different colors represent SNs with different final atmosphere fractions. (a)--(c) $N_\text{ini} = 60, 80,$ and 100 and $e_\text{ini} = 0.9$. (d)--(f) $N_\text{ini} = 60, 80,$ and 100 and $e_\text{ini} = 0.8$. (g)--(i) $N_\text{ini} = 60, 80,$ and 100 and $e_\text{ini} = 0.7$. The green, blue, and gray dashed lines represent the observed peak for super-Earths (1.5 $R_\oplus$), the position of the radius valley, and the peak for SNs (2.7 $R_\oplus$), respectively. The color indicates the atmospheric percentages (see color bar).
\label{fig:fig4}}
\end{figure*}

Here we summarize how the size distribution of SNs changes as a result of SN--embryo collisions. Figure \ref{fig:fig4} summarizes the initial and final size distributions for 20 runs of simulations with different numbers of embryos and eccentricities. 

\begin{figure*}[ht!]
\plotone{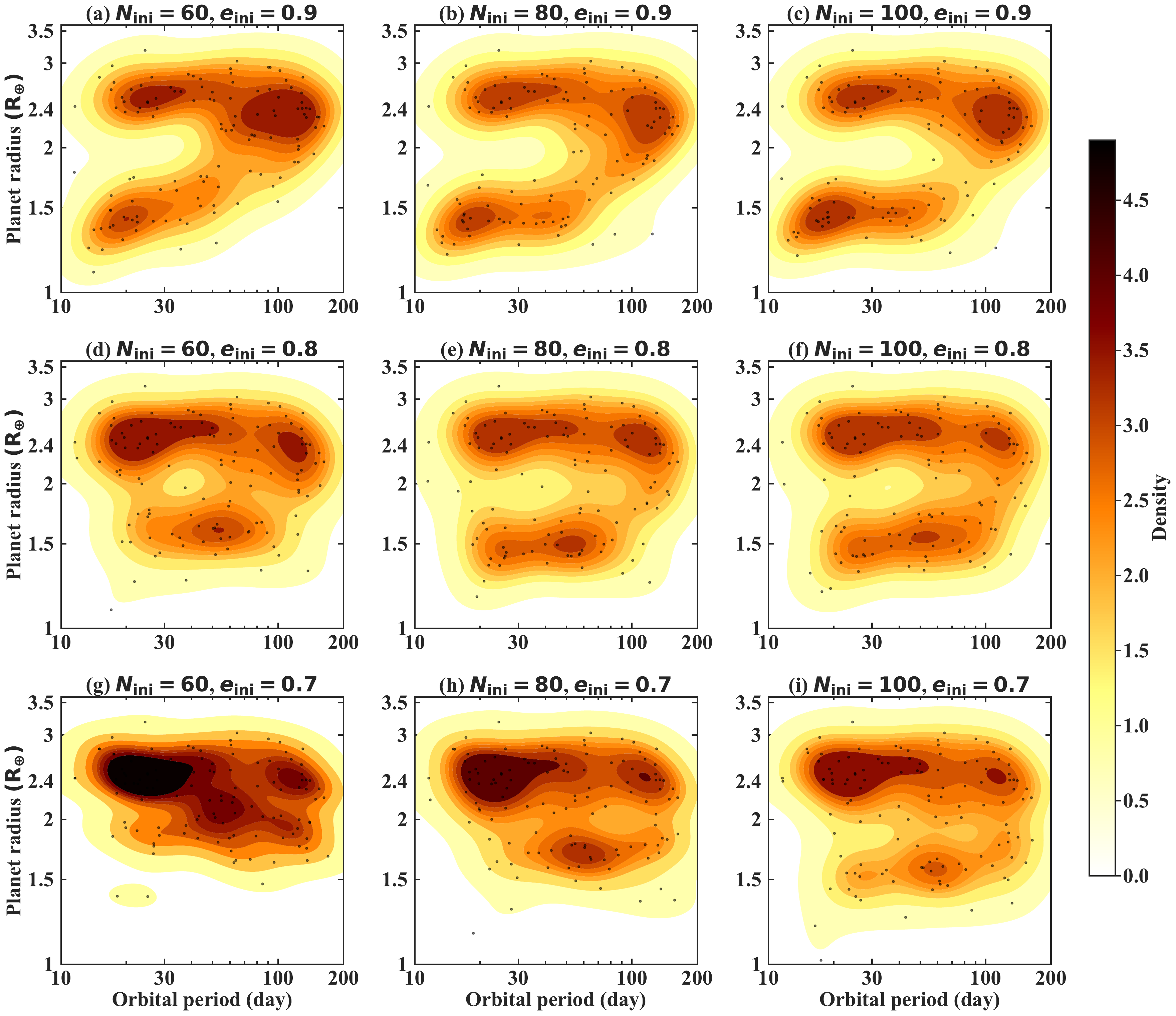}
\caption{The same as Figure \ref{fig:fig4}, but showing the occurrence of planets by combining the initial and final distributions in the orbital-period--planet-radius plane. Contours in these plots connect points of equal probability density. The densities shown, estimated using kernel-density estimation, represent statistical probability densities that indicate the likelihood of finding planets with specific combinations of orbital period and radius. The integral of the density over any region gives the probability of finding planets within that area. The unit area is defined in logarithmic space, where the units are $\log_{10}$(days) for the orbital period and $\log_{10}$($R_{\oplus}$) for the planet radius. The density is therefore expressed in units of 1/[$\log_{10}$(days) $\times$ $\log_{10}$($R_{\oplus}$)], and the integral over the entire space is normalized to unity. 
\label{fig:fig5}}
\end{figure*}

As we have seen, collisions between SNs and embryos can result in atmospheric mass loss and changes in the size distribution of the SNs.
For $e_\text{ini} = 0.8$ and $N_\text{ini} = 80$ [Figure \ref{fig:fig4}(e)], corresponding to the case shown in Figure \ref{fig:fig1}, there are 13 SN--embryo collisions per SN on average (Table~\ref{tab:tab1}). At the end of the simulation, the size distribution has a peak at 1.5 $R_{\oplus}$, with an atmospheric fraction $<0.1\%$. A bimodal distribution with a radius gap is thus formed in this case. While most SNs lose the majority of their atmosphere, 32\% retain more than 10\% of their initial atmosphere (i.e., $f_{\rm atm} \gtrsim 0.3\%$) (Table \ref{tab:tab1}). Consequently, this mechanism can also account for the formation of planets residing within the radius gap.

Next, we examine how the size distribution changes when the parameters are varied. As noted above, the atmospheric mass loss is more pronounced for cases with larger initial eccentricities and larger initial numbers of embryos, which is also seen in Figure \ref{fig:fig4}. 
A bimodal radius distribution with distinct peaks at 1.5 and 2.7 $R_{\oplus}$ is observed in systems with higher initial eccentricities and larger numbers of embryos. For $e_\text{ini} = 0.9$, this bimodality appears in systems with 60, 80, and 100 embryos [Figures \ref{fig:fig4}(a)--(c)]. At $e_\text{ini} = 0.8$, it persists in systems with 80 and 100 embryos [Figures \ref{fig:fig4}(e) and (f)], while for $e_\text{ini} = 0.7$, it is evident only in the 100 embryo case [Figure \ref{fig:fig4}(i)]. In these configurations, each SN experiences approximately 13 collisions on average. Notably, the bimodal distribution becomes prominent when $ F_{X_{\text{total}} < 0.9} $ is less than about 40\% (Table \ref{tab:tab1}).

On the other hand, for $e_\text{ini} = 0.7$ and $N_\text{ini} = 80$ [Figure \ref{fig:fig4}(h)], there are 10 SN--embryo collisions per SN (see also Table~\ref{tab:tab1}). At the end of the simulation, the size distribution peaks at 1.8 $R_{\oplus}$ (corresponding to 0.7\% of the atmospheric mass fraction). This is slightly larger than the observed peak at 1.5 $R_\oplus$. The number of SNs that retain more than 10\% of the initial atmosphere is 77\% (Table \ref{tab:tab1}), and the majority of SNs retain about 0.1\% or more of the atmosphere at the end. Although in this case we cannot reproduce the observed locations of the peaks and the gap, planets can still form inside the location of the observed radius gap.

These results demonstrate that planets can indeed fall into the radius gap. As shown in Figure \ref{fig:fig1}(c), the atmospheric mass lost in each collision is the product of a factor $X$ and the atmospheric mass fraction $f_{\rm atm}$ at the time of impact. Significant atmospheric mass loss therefore occurs during the first few collisions, when the atmospheric mass fraction is still large. Consequently, the evolution from SNs with an initial atmosphere of $\simeq$3\% to gap planets with $\simeq$0.3\%--1\% occurs relatively easily, as the atmospheric fraction drops below this target range after about 3--6 collisions. In fact, in all the simulations shown in Figure~\ref{fig:fig4}, some planets form with a size corresponding to the gap planets, around $2\,R_\oplus$.

We also examine the evolution of SN eccentricities. Recent observational studies indicate that among SNs with radii ranging from 1 to 3 $R_\oplus$, those residing within the radius gap tend to display higher eccentricities \citep{2025PNAS..12205295G}. In our simulations, the initial eccentricity of the SNs is set to 0.02, consistent with the observed values of many SENs. Due to close encounters between planetary embryos and SNs, the eccentricities of the SNs increase to about 0.05 or higher, which may explain the observed trend. Note that our results show that SNs with radii of $\sim 1.5 R_\oplus$ also exhibit increased eccentricities. Therefore, either the eccentricities of planets with $R<2R_\oplus$ must be damped afterward, or the majority of such planets may need to be explained by mechanisms other than impact erosion.

Our results may also explain the origin of the radius cliff: as planets lose their atmospheres through these interactions, their radii shrink, contributing to the sharp drop in occurrence observed between 2.5 and 4 $R_\oplus$. 
Although in our baseline setup we adopt an initial mean atmospheric mass fraction of $\simeq 3\%$, SNs may also accrete more massive atmospheres, such as $\sim 10\%$. In Section~\ref{sec:sec4.3}, we examine how the planetary radius evolves in selected cases when the initial atmospheric mass fraction is larger. These results suggest that impact erosion may contribute to the formation of the radius cliff.

For cases where we see a clear gap in the 1D size distribution, we also see a gap in the 2D plots. Figure \ref{fig:fig5} shows how the occurrence of planets depends on the orbital period. 
First, we note that a clear gap exists at the location corresponding to a radius of $2\,R_\oplus$.
Panels (a), (b), (c), (e), (f), and (i) in this figure show that a radius gap exists around 2 $R_\oplus$, located between the two peaks at 1.5 $R_\oplus$ and 2.7 $R_\oplus$, as shown in Figure \ref{fig:fig4}.
Next, we examine how the orbital locations showing a deficit in occurrence relate to the initial setup. In Figures \ref{fig:fig5}(a)--(c), we observe that SNs with significant radius reductions typically have the shortest periods ($\sim$ 10--30 days), while Figures \ref{fig:fig5}(d)--(f) indicate that SNs at periods $\sim$ 30--60 days experience significant mass loss. Figures \ref{fig:fig5}(g)--(i) show that SNs with radius reductions have periods of around 60--100 days.

Note that the profile of the gap and its radial dependence are rather complicated. Figure~\ref{fig:fig5} combines the initial and final size distributions of planets in the radius--period plane, in an analogous way to Figure~\ref{fig:fig4}. 
Embryos on high-eccentricity orbits undergo multiple collisions with SNs, efficiently stripping their atmospheres and producing SEs, as reflected in the final radius distributions for systems with large $e_{\mathrm{ini}}$ and $N_{\mathrm{ini}}$. If the eccentricities of embryos are low or if the number of embryos is small, collisions between embryos and SNs are rare, so planets largely retain their initial atmospheres and remain as SNs; in this regime, the population is effectively described by the initial radius distribution.
In the radius--period plane, the slope of the gap is positive for $e_\text{ini} = 0.9$; it becomes nearly flat when $e_\text{ini}$ decreases to $0.8$ and may even become negative for $e_\text{ini} = 0.7$.
Note, however, that this may be slightly affected by our use of test-particle simulations. If instead the mutual gravitational interactions among the embryos were considered, it would likely produce a more-uniform distribution of eccentricities among the embryos, as well as a more-even frequency of collisions across different orbital periods; see also Section~\ref{sec:sec4.1} below. On the other hand, depending on the conditions, it is also possible for outer SNs to lose significant atmospheric mass, as do inner SNs (e.g., $e_\text{ini} = 0.8$, $N_\text{ini} = 80$). As a result, the radial dependence of the gap profile is different. For $e_\text{ini} = 0.9$, the most significant loss of radius by SNs occurs within the 10--30 day range; for $e_\text{ini} = 0.8$, it is between 30--100 days; and for $e_\text{ini} = 0.7$, between 60--170 days. In summary, the distribution on the 2D map depends on the eccentricities and numbers of the embryos. We may constrain the properties of the remaining high-eccentricity embryos by comparison with observations. Additional simulations should be performed, and the resulting gap slopes and their radial dependence should be compared with those observed in SEs and SNs in a future study.

\section{Discussion} \label{sec:Discussion}
\subsection{Results of full \textit{N}-body simulations}\label{sec:sec4.1}
\begin{figure*}[ht!]
\plotone{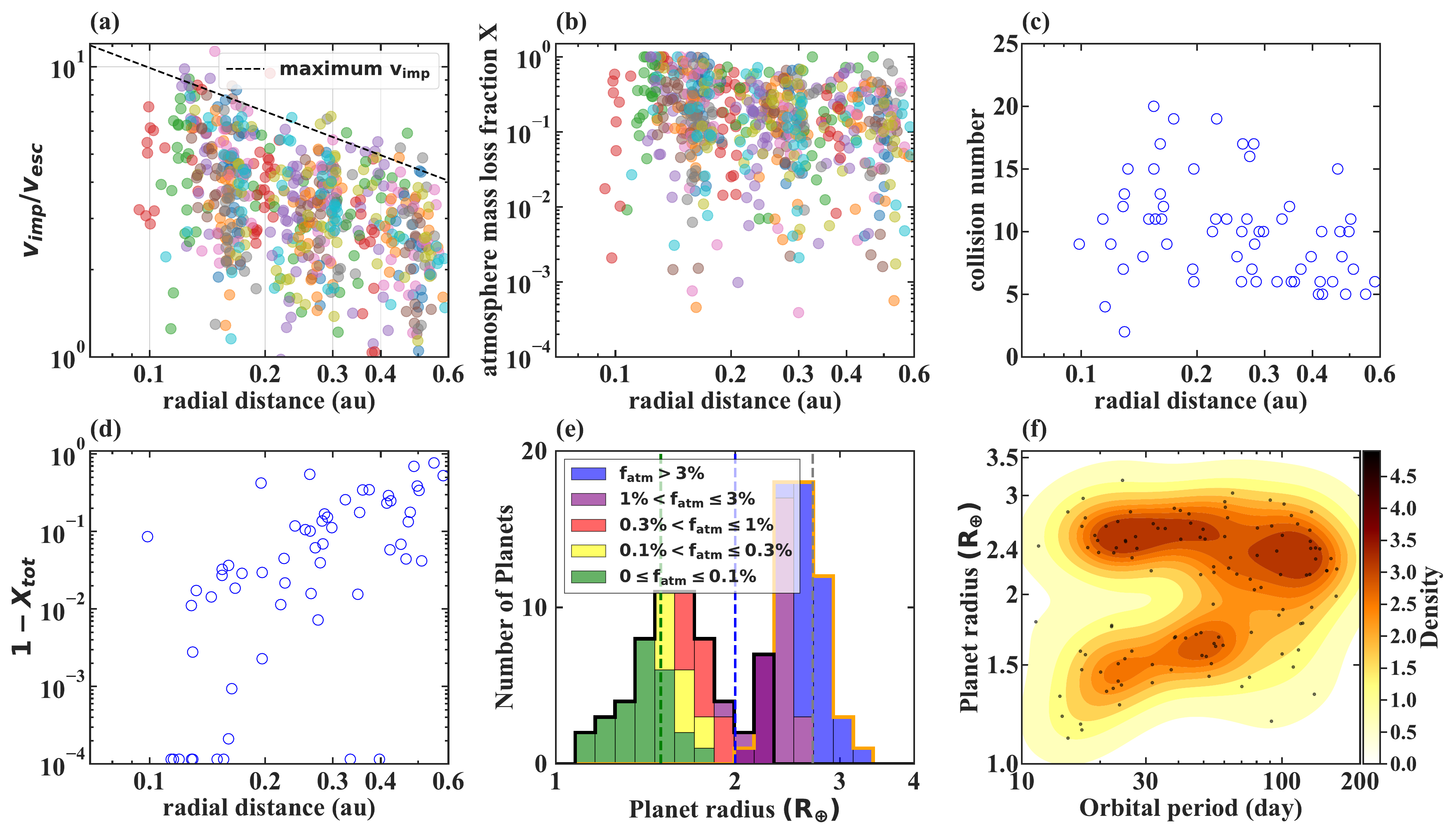}
\caption{Results of full \textit{N}-body simulations with the initial conditions $e_\text{ini} = 0.8$ and $N_\text{ini} = 80$. (a) Impact velocities between SNs and embryos. (b) Atmospheric mass-loss fraction. (c) Total number of collisions. (d) Final remaining atmospheric mass fractions of SNs. (e) Final radius distribution of SNs at 50 Myr and the initial radius distribution. (f) Dependence of the final radius of SNs at 50 Myr and initial radius on orbital-period distribution. 
\label{fig:fig6}}
\end{figure*}

In section \ref{sec:results}, we showed the results of simulations that do not take into account the mutual gravity between the embryos, as such test-particle simulations can save computational resources. However, we have also performed full \textit{N}-body simulations with mutual gravitational interactions among the embryos.
 The initial conditions are identical to those in Section \ref{sec:results}, with $N_\text{ini} = 80$ and $e_\text{ini} = 0.8$ . Our purpose in performing these full \textit{N}-body simulations was to compare the results with those of the test-particle simulations, in order to ensure that the results of the test-particle simulations presented in Section \ref{sec:results} are reliable.
 Figure \ref{fig:fig6} shows the results for these simulations, which can be compared with Figures~\ref{fig:fig2}--\ref{fig:fig5}.
Some statistics are also shown in the last row of Table \ref{tab:tab1}.

Here, we assess whether the findings in Section \ref{sec:results} hold in the full \textit{N}-body simulations. Both the test-particle and full \textit{N}-body simulations reveal frequent high-velocity collisions between SNs and embryos, which lead to substantial atmospheric stripping [Figures \ref{fig:fig2}(b) and (e), and \ref{fig:fig6}(a) and (b)]. The full \textit{N}-body simulations, however, exhibit several differences. For example, embryos attain higher maximum eccentricities via mutual scattering, resulting in slightly larger maximum impact velocities (Table \ref{tab:tab1}). Although the atmospheric mass loss per collision is increased, the total loss fraction remains similar to that in the test-particle case because the overall number of SN--embryo collisions is reduced (Table \ref{tab:tab1}). In addition, while test-particle simulations show a strong radial gradient in the collision frequency [Figure \ref{fig:fig3}(b)], the broader eccentricity distribution in the full \textit{N}-body runs weakens this dependence [Figure \ref{fig:fig6}(c)], allowing collisions even at $\sim$0.1 au [Figure \ref{fig:fig6}(a)]. Some SNs beyond 0.4 au still undergo nearly complete ($\simeq$100\%) atmospheric loss [Figures \ref{fig:fig6}(b) and (d)]. Despite these differences, both methods yield broadly consistent outcomes: they reproduce similar radius distributions, recover the observed radius gap [Figures \ref{fig:fig4}(e) and \ref{fig:fig6}(e)], and produce comparable radius--period distributions, including collisions involving SNs with orbital periods longer than 100 days [Figures \ref{fig:fig5}(e) and \ref{fig:fig6}(f)].

\subsection{Validity of initial conditions}\label{sec:sec4.2}
\subsubsection{Formation of high-eccentricity embryos}
Here, we discuss the plausibility of the existence of the high-eccentricity embryos assumed in this study.
Previous \textit{N}-body simulations have suggested that high-eccentricity embryos can remain in orbits around 1 au after the formation of SNs \citep{2017A&A...607A..67M,2020ApJ...899...91O,2024arXiv240410831I}. However, in previous \textit{N}-body simulations of solar-system formation, high-eccentricity embryos do not remain after the formation of the terrestrial planets. Accordingly, we present several possible explanations for the formation of such high-eccentricity embryos during the formation of SNs.

The escape eccentricity provides a rough estimate of the eccentricity after a close encounter between SNs and embryos \citep{1972IAUS...45..329S,2013ApJ...775...42I,2020A&A...642A..23M}:
\begin{eqnarray}
e_\text{esc} = \frac{v_{\text{esc}}}{v_{K}}&=& 0.3\left(\frac{M_{\text{core}} + M_{\text{emb}}}{M_{\oplus}}\right)^{\frac{1}{3}}\left(\frac{\rho}{4.5\,\text{g}\,\text{cm}^{-3}}\right)^{\frac{1}{6}}\nonumber\\
&&\times\left(\frac{r}{1\,\text{au}}\right)^{\frac{1}{2}}.
\end{eqnarray}
The eccentricity is larger for larger masses of SNs and larger orbital distances. In previous \textit{N}-body simulations of solar system formation, the mass of the core is assumed to be up to 0.1 $M_{\oplus}$, so the escape eccentricity is about 0.1 \citep[e.g.,][]{2012PTEP.2012aA308K}. However, in the case of SN formation, the mass is 30 times larger, and the escape eccentricity becomes about 0.4. Of course, this is only a rough estimate. Therefore, some previous studies use a Rayleigh distribution with the scale parameter of $e_\text{esc}$ to formulate the eccentricity distribution, which is in agreement with \textit{N}-body simulations \citep[e.g.,][]{2013ApJ...775...42I}.

The escape eccentricity given above does not include a dependence on the mass ratio between the SNs and embryos. However, the eccentricities of the embryos are expected to depend on this mass ratio. By considering angular-momentum conservation before and after one close encounter between an SN and an embryo \citep{2013ApJ...775...42I}, we obtain
\begin{eqnarray}
&&M_\text{core}\sqrt{a_\text{core,0}} 
+ M_\text{emb}\sqrt{a_\text{emb,0}}\nonumber\\ &=&M_\text{core}\sqrt{a_\text{core}(1+e_\text{core})(1-e_\text{core})} \nonumber \\
&+& M_\text{emb}\sqrt{a_\text{emb}(1-e_\text{emb})(1+e_\text{emb})}\\
&=& M_\text{core}\sqrt{a_\text{core,0}(1-e_\text{core})}\nonumber \\
&+&M_\text{emb}\sqrt{a_\text{emb,0}(1+e_\text{emb})},
\end{eqnarray}
where $a_{\text{core},0}$ and $a_{\text{emb},0}$ are the semimajor axes of the SN and embryo before the close encounter, and $a_{\text{core}}$, $e_{\text{core}}$, $a_{\text{emb}}$, and $e_{\text{emb}}$ are the semimajor axes and eccentricities of the SN and embryo just after the close encounter. We assume $a_{\text{core},0}\approx a_\text{core}(1+e_\text{core})$ and $a_{\text{emb},0}\approx a_\text{emb}(1-e_\text{emb})$. If $e_\text{core}$ is a small quantity, then using $\sqrt{1-e_\text{core}}\approx 1-e_\text{core}/2$, we have
\begin{eqnarray}
&&M_\text{core}\sqrt{a_{\text{core},0}} + M_\text{emb}\sqrt{a_{\text{emb},0}}
\nonumber\\&=& M_\text{core}\sqrt{a_{\text{core},0}} \left(1 - \frac{1}{2}e_\text{core}\right) \nonumber\\
&+& M_\text{emb}\sqrt{a_{\text{emb},0}(1 + e_\text{emb})}.
\end{eqnarray}
Solving this for $e_\text{emb}$, we get
\begin{eqnarray}
&&e_\text{emb} = \left(\frac{e_\text{core}}{2}\frac{M_\text{core}}{M_\text{emb}}\sqrt{\frac{a_{\text{core},0}}{a_{\text{emb},0}}} + 1\right)^{2} - 1\\
&=& e_\text{core}\frac{M_\text{core}}{M_\text{emb}}\sqrt{\frac{a_{\text{core},0}}{a_{\text{emb},0}}} + \frac{e_\text{core}^{2}}{4}\left(\frac{M_\text{core}}{M_\text{emb}}\right)^{2}\frac{a_{\text{core},0}}{a_{\text{emb},0}}.
\end{eqnarray}
This indicates that when the mass ratio between SNs and embryos is large, the eccentricities of the embryos can be excited significantly. If we assume $M_{\text{core}}=3M_\oplus$ and $M_{\text{emb}}=0.05M_\oplus$, then the mass ratio is 60. Therefore, depending on the eccentricity of the SNs and the semimajor axes of the SNs and the embryos, the eccentricities of the embryos can be quite high; e.g., for $e_\text{core}=0.02$, $e_\text{emb}$ can be unity. In some previous \textit{N}-body simulations of SN formation, most embryos have been large ($>0.1M_\oplus$), and this effect is not clearly observed. Even with smaller embryos ($<0.1M_\oplus$), this point has not drawn sufficient attention.

We have also done test simulations for the scattering of embryos by SNs during formation. As discussed in Section \ref{sec:intro}, we consider a case where SNs form at about $a=1$ au and slowly undergo inward migration. We have performed two different simplified simulations of this process. First, we performed a simulation with a single SN of 3 $M_\oplus$ at $a_\mathrm{SN}=0.8$ au, surrounded by 200 planetary embryos, each with a mass of 0.05 $M_\oplus$, distributed between $a=0.63$--0.97 au. In another simulation, two SNs are placed at $a_\mathrm{SN}=0.6$ au and $a_\mathrm{SN}=0.8$ au, with 200 embryos similarly distributed in the range $a=0.44$--0.97 au. In both setups, the initial eccentricities of the SNs and embryos are set to 0.02. The simulations demonstrate that scattering caused by the SNs can indeed excite the eccentricities of the embryos to large values ($e>0.4$). In the case with a single SN, the frequency of strong scattering events is limited because the gravitational interactions between the embryos and the SN become weak when they are scattered to positions farther from the SN, and the maximum eccentricity of the embryos reaches $\sim$ 0.6 [see also Figure \ref{fig:figA2}(a)]. However, in the case with two SNs, the additional scattering from the inner SN further amplifies the eccentricities of the embryos in outer orbits, leading to even greater eccentricities [Figure \ref{fig:figA2}(b)]. For further details, see Appendix \ref{sec:appendix}.

\subsubsection{Eccentricity-damping timescale}
After high-eccentricity embryos form as a byproduct of SN formation, they can remain at high eccentricity if the eccentricity is not significantly damped during the existence of the protoplanetary gas disk. We therefore estimate the timescale for eccentricity damping by disk gas. The timescale for eccentricity damping by planet--disk interactions is given by \citep[e.g.,][]{2008A&A...482..677C,2017MNRAS.470.1750I}
\begin{eqnarray}
t_{e} &=& \frac{t_{\text{wave}}}{0.78}\left[1 - 0.14\left(\frac{e}{h}\right)^{2} + 0.06\left(\frac{e}{h}\right)^{3}+ 0.18\left(\frac{e}{h}\right)\right.\nonumber \\
      && \left.\times \left(\frac{i}{h}\right)^{2}\right]\label{eq:t_e}
      \\
      &=& 8.5\,\text{kyr}\left(\frac{M_{\text{emb}}}{0.05M_{\oplus}}\right)^{-1}\left(\frac{r}{1\,\text{au}}\right)^{2}\left(\frac{L_*}{L_\odot}\right)^{\frac{1}{2}}\nonumber \\
      &&\times\left[1 - 0.14\left(\frac{e}{h}\right)^{2}\right.
       \left. + 0.06\left(\frac{e}{h}\right)^{3} +0.18\left(\frac{e}{h}\right)\right.\nonumber \\
     &&\times\left.\left(\frac{i}{h}\right)^{2}\right]\label{eq:t_e2},
      \end{eqnarray}
where
\begin{eqnarray}
t_{\text{wave}} &=& \left(\frac{M_{\odot}}{M_{\text{emb}}}\right)\left(\frac{M_{\odot}}{\Sigma_{\text{gas}}a^{2}}\right) h^{4}\Omega_{K}^{-1}.
\end{eqnarray}
The following properties of the disk are considered.
\begin{eqnarray}
\Sigma_{\text{gas}} &=& 1700\left(\frac{r}{1\,\text{au}}\right)^{-\frac{3}{2}}\,\text{g}\,\text{cm}^{-2}, \\
c_{s} &=& \sqrt{\frac{k_{b}T}{\mu m_{H}}}, \\
T &=& 280\left(\frac{r}{1\,\text{au}}\right)^{-\frac{1}{2}}\left(\frac{L_*}{L_{\odot}}\right)^{\frac{1}{4}}\,\text{K}, \\
h &=& \frac{c_{s}}{v_{K}} = 0.033\left(\frac{r}{1\,\text{au}}\right)^{\frac{1}{4}}\left(\frac{L_*}{L_{\odot}}\right)^{\frac{1}{8}},\label{eq:damping}
\end{eqnarray}
where $h$, $\Sigma_{\text{gas}}$, $\Omega_{K}$, $\mu$, $m_{H}$, and $L_*$ are the disk scale height, gas surface density, the Keplerian orbital frequency of the planet, the mean molecular weight, the mass of a hydrogen atom, and the stellar luminosity, respectively. The farther the planet is from the star and the smaller the masses of the planetary embryos, the longer is the damping timescale. In addition, at high eccentricities the correction factor in Equations~(\ref{eq:t_e})--(\ref{eq:t_e2}) increases significantly; for example, at $e=0.5$ the damping timescale is longer by a factor of about 100--1000. For planetary embryos with mass $0.05\,M_{\oplus}$, eccentricity $e=0.8$, and inclination $i=0.8 \,\text{rad}$, located at distances in the range $r=1$--2 au, the damping timescale ranges from 50 Myr to 100 Myr, which is significantly longer than the timescale for dissipation of the gas disk. Therefore, the high eccentricities of the embryos can still remain after the gas disk dissipates.

\subsubsection{Accretion timescale}
The collision timescale between embryos must also be sufficiently long for high-eccentricity embryos to remain after their formation.
We take the accretion rate between embryos to be \citep{1993Icar..106..210I} 
\begin{eqnarray}
\dot{M}_{\text{acc, em}} &=& \Sigma_{\text{solid}} \Omega_{\text{K}} \pi R^2 \left( 1 + \frac{v_{\text{esc}}^2}{\nu^2} \right),
\end{eqnarray}
where $\Sigma_{\text{solid}}$ is the surface density of solid particles and $\nu$ is the velocity dispersion.  The accretion timescale between emrbyos thus becomes
\begin{eqnarray}
t_{\text{acc}} &=&\frac{M_{\text{emb}}}{\dot{M}_{\text{acc, em}}} \\
&\approx& 6 \times 10^{6}\,\text{yr}\left(\frac{\Sigma_{\text{solid}}}{77\,\text{g}\,\text{cm}^{-2}}\right)^{-1}  \left(\frac{a}{1\,\text{au}}\right)^{\frac{3}{2}} \nonumber \\\label{t_acc}
&& \times \left(\frac{M_{\text{emb}}}{M_{\oplus}}\right)^{\frac{1}{3}}  \left(1 + \frac{v_{\text{esc}}^{2}}{\nu^{2}}\right)^{-1}\\
&\approx& 1.4 \times 10^{7}\,\text{yr}\left(\frac{a}{1\,\text{au}}\right)^{\frac{3}{2}}  \left(\frac{M_{\text{emb}}}{0.05M_{\oplus}}\right)^{\frac{1}{3}}.
\end{eqnarray}
This indicates that the accretion timescale between embryos is 14 Myr for embryos with $M_{\rm emb} =0.05 \,M_\oplus$ at $a = 1$  au. The accretion timescale is larger than the dissipation timescale of the gas disk, so a population of low-mass embryos can survive after disk dispersal. Therefore, high-eccentricity embryos are not depleted rapidly through collisions among embryos. Note that we could also estimate the collision timescale between SNs and embryos using the same method. However, to apply it to collisions between two populations with different semimajor axes, we would need to include some correction factors in the formula. Instead, we read the SN--embryo collision timescale directly from Figure~\ref{fig:fig1}. From Figure~\ref{fig:fig1}, the first SN--embryo collision occurs at about 1 Myr. The median time of SN--embryo collisions is about 5 Myr.

\begin{figure*}[ht!]
\plotone{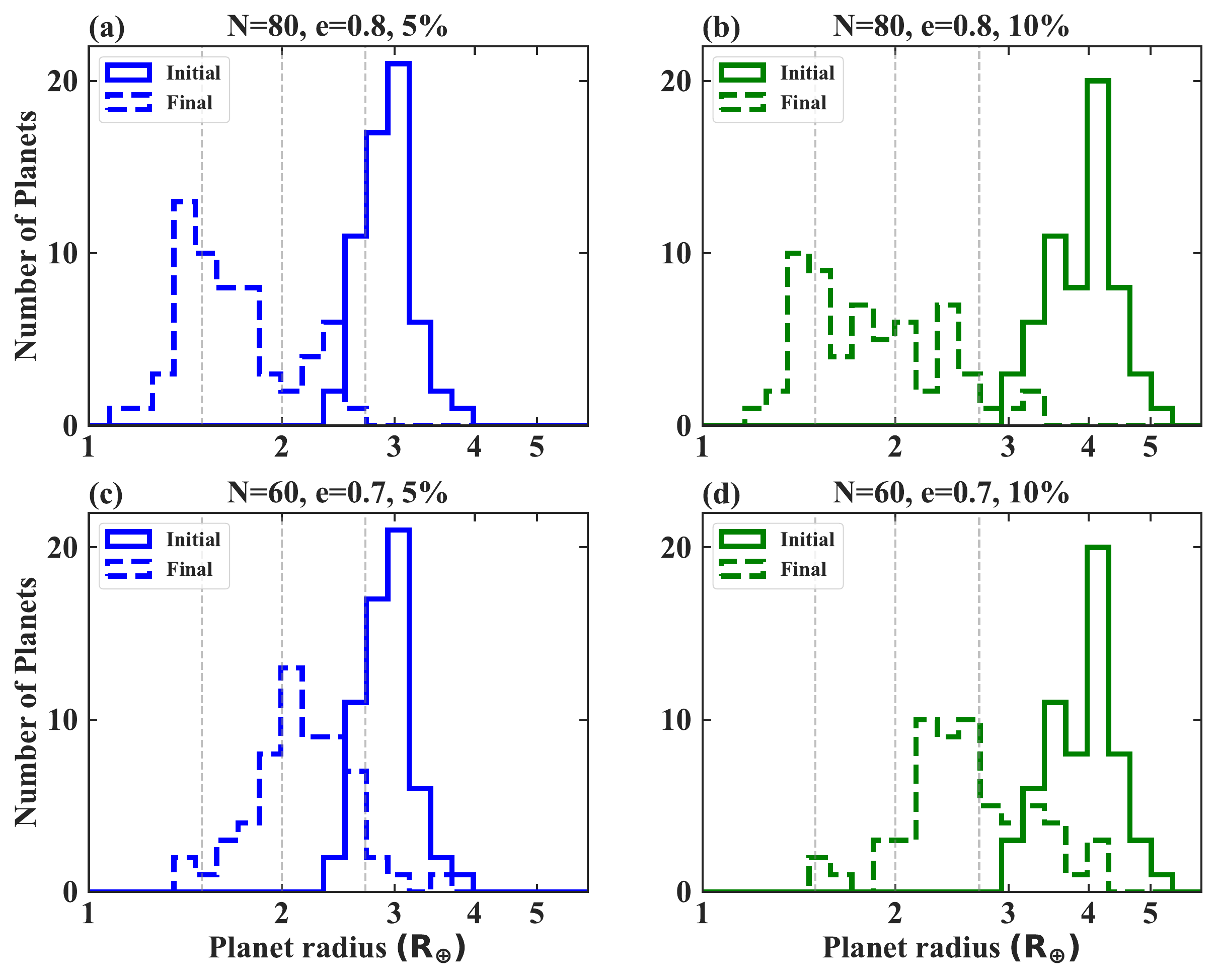}
\caption{Initial and final radius distributions for different assumed initial atmospheric mass fractions. Panels (a) and (b) present results for the fiducial case ($N_{\mathrm{ini}}=80$, $e_{\mathrm{ini}}=0.8$) with mean initial atmospheric mass fractions of 5\% and 10\%. Panels (c) and (d) show results for the $N_{\mathrm{ini}}=60$, $e_{\mathrm{ini}}=0.7$ case with the same two mass fractions.
\label{fig:fig7}}
\end{figure*}

\subsection{Dependence on the initial atmospheric mass fraction}\label{sec:sec4.3}
In this work, for the initial atmospheric mass fractions of SNs, we use a log-normal distribution centered at 3\% with a standard deviation of 0.3, a choice that is analogous to previous studies \citep[e.g.,][]{2021MNRAS.503.1526R}. However, formation models allow SNs to acquire a wider range of atmospheric masses, depending on disk properties and accretion histories \citep[e.g.,][]{2014ApJ...797...95L,2015ApJ...811...41L,2016ApJ...825...29G}. In addition, the initial atmospheric mass fraction may also vary with planet mass and orbital period \citep[e.g.,][]{2017ApJ...847...29O}. To assess sensitivity to this assumption, we perform additional post-processing calculations in which the initial atmospheric mass fraction is set to mean values of 5\% and 10\% for two cases.

Figure~\ref{fig:fig7}(a)–(b) show the initial and final radius distributions for the fiducial model ($N_{\rm ini}=80$, $e_{\rm ini}=0.8$). We find that even when the initial atmospheric mass fractions are increased to 5\% or 10\%, the fiducial model still produces a radius gap. In both cases, the final radius distribution peaks at $1.5\,R_{\oplus}$, consistent with the 3\% baseline result shown in Figure~\ref{fig:fig4}(e). This is because each SN undergoes on average about 13 collisions with planetary embryos, and each collision removes $\sim 20\%$ of the remaining atmosphere (Table~\ref{tab:tab1}). Hence, in systems with high collision numbers, impact erosion produces a radius gap whether the initial atmospheric fractions of the planets are 3\%, 5\%, or 10\%.

We then examine the cases in which the number of collisions is minimized ($N_{\rm ini} = 60$, $e_{\rm ini} = 0.7$). When the initial mean atmospheric mass fraction is 5\%, the final radius distribution is concentrated at $\simeq 2\,R_{\oplus}$ (Figure~\ref{fig:fig7}(c)), within the observed radius gap, consistent with the result in Figure~\ref{fig:fig4}(g). This outcome indicates that a combination of low initial atmospheric mass fractions (3--5\%) and low collision frequencies yields planets residing inside the gap. 
For an initial mean atmospheric mass fraction of 10\% (Figure~\ref{fig:fig7}(d)), we still find that some planets remain within the radius gap. These results also provide implications for the formation of the radius cliff. Even when SNs initially accrete relatively massive atmospheres of 5\% or 10\%, such that their initial radii exceed $3\,R_\oplus$, impact erosion can reduce their radii. Indeed, Figure~\ref{fig:fig7}(c) and (d) show that, even in our minimum collision case, the occurrence of SNs with radii larger than $2.5\,R_\oplus$ declines.
A detailed assessment of these effects will require further dedicated study.

\subsection{Caveats of the planetary structure model}\label{sec:radius_model_limits}
For computing planetary radii, analytic fits from \citet{2014ApJ...792....1L} are often used. In their model, a convective H/He envelope lies beneath an isothermal, radiative atmosphere whose thickness is approximated by Equation~(\ref{eq:radius}).
Recent work, however, shows that early-time boil-off can modify the core-envelope cooling history in ways not captured by these simplified prescriptions. As emphasized by \citet{2024MNRAS.528.1615O} and \citet{2024ApJ...976..221T}, the feedback between thermal evolution and atmospheric escape may be important. \citet{2025ApJ...989...28T} further demonstrates that rapid mass loss leads to adiabatic expansion of the envelope, which increases the temperature contrast between the rock core and the base of the envelope, enhances the core heat flux, and sustains higher post-boil-off envelope entropy. As a result, planets can retain somewhat larger radii at fixed H/He mass fraction than predicted by models such as \citet{2014ApJ...792....1L} that neglect this coupling. In addition, simplified radiative-atmosphere prescriptions can underestimate the radii of low-mass planets \citep[e.g.,][]{2025ApJ...989...28T}.

As described in Section~\ref{sec:post}, we compute the planet size as $R_{\rm SN} = R_{\rm core} + R_{\rm env}$, omitting the radiative-atmosphere term $R_{\rm atm}$ to remain consistent with the one-dimensional structure calculations of \citet{2014ApJ...792....1L}.
A more accurate calculation of planetary radii using next-generation structure models \citep{2025ApJ...989...28T} will be pursued in future work.

\section{Conclusions} \label{sec:Conclusions}
The radius gap observed in SNs, characterized by a deficit in the occurrence of planets between approximately 1.5 and 2.7 $R_\oplus$, has attracted considerable attention in recent exoplanet studies, and previous research has attributed this feature to atmospheric escape mechanisms such as photoevaporation and core-powered mass loss. However, observations also reveal that some planets have radii within the radius gap. The present study aims to quantify the impact velocities between close-in SNs and outer, high-eccentricity planetary embryos, exploring how these collisions contribute to atmospheric loss and influence the final radius distribution of SNs, particularly for planets inside and outside the observed radius gap. Utilizing \textit{N}-body simulations that extend over 50 Myr, we have modeled the dynamics of planetary systems containing three close-in SNs and outer-region planetary embryos with various initial eccentricities and numbers of embryos, to investigate the effects of initial conditions on collision outcomes and atmospheric erosion.

Our main findings are as follows:
\begin{itemize}
  \item Outer high-eccentricity embryos collide with inner SNs on timescales of several million years. These collisions occur at velocities 2--5 times the escape velocity, with each impact removing 15\%--30\% of the SN atmosphere. After 3--6 such collisions, the atmospheric mass decreases to about 1/3 of its initial amount.
  \item The atmospheric mass-loss fraction depends on the impact velocity; thus higher atmospheric mass-loss rates occur in simulations with higher initial eccentricities ($e_\text{ini} = 0.8$, 0.9) and for SNs at smaller orbital distances.
  \item Across all our simulations, SNs undergo, on average, more than six collisions. Consequently, SNs that initially retain atmospheres with $\sim$ 3\% of the planetary mass end up with atmospheric mass fractions below 1\%. Since planets with atmospheres containing 0.3\%--1\% of the mass fall within the radius gap, this accounts naturally for the existence of such planets.
  \item For systems with $N_\text{ini} = 80$ and high eccentricities ($e_\text{ini} = 0.8$, 0.9), SNs undergo > 13 collisions, and most of these SN atmospheres are dissipated, resulting in the emergence of a clear radius gap.
 \item These findings remain unchanged even when mutual interactions between embryos are considered. 
 \item In systems where SNs initially have relatively massive primordial atmospheres (e.g., $\sim$10\%), collisions can reduce the atmospheric fraction to $\sim$3\% even for our minimum collision case. This result provides implications for the origin of the radius cliff.
\end{itemize}

Our simulations show that impact erosion can also operate beyond 100-day orbital periods. This distinguishes our mechanism from photoevaporation, which becomes less effective at larger orbital distances. This prediction can be tested with future observations. Another key distinction between our mechanism and photoevaporation lies in the formation of gap planets. In the photoevaporation scenario, SNs that begin to lose their atmospheres are unlikely to survive within the radius gap. In contrast, under our impact-erosion model, it is relatively easy for SNs to lose $\sim$2/3 of their initial atmospheric mass, but further loss becomes increasingly limited. This readily explains the origin of planets within the radius gap. Recent observations also suggest that planets within the gap may have slightly elevated eccentricities, and our model potentially explains this trend, as the presence of high-eccentricity embryos tends to increase the eccentricities of SNs. This will be studied in future work.

\begin{acknowledgments}
We thank Hiroyuki Kurokawa for his comments, which helped improve the paper. We also thank Yuji Matsumoto for valuable discussions. We thank the anonymous referee for a careful and constructive review, which significantly improved the clarity of the manuscript. This work is supported by the National
Natural Science Foundation of China (grant No. 12273023). Kangrou Guo is supported by the National Natural Science Foundation of China (grant No. 12503069) and acknowledges the support from the K. C. Wong Educational Foundation.
\end{acknowledgments}

\appendix 
\section{Scattering simulation of embryos by SNs}\label{sec:appendix}
 To demonstrate how the eccentricities of embryos are excited by SNs during formation, we perform test simulations. We set two different sets of initial conditions. The first set includes a single SN of 3 $M_{\oplus}$ at $a_\mathrm{SN}=0.8$\,au, with 200 embryos placed at $a=0.63$--$0.97$\,au. The second set involves one SN at $a_\mathrm{SN}=0.6$\,au and another at $a_\mathrm{SN}=0.8$\,au with 200 embryos placed at $a=0.44$--$0.97$\,au. All embryos have a mass of $0.05\,M_{\oplus}$. The initial eccentricities for both the SNs and the embryos are $0.02$. We conducted test particle simulations without considering interactions between embryos.

\begin{figure*}[ht!]
\plotone{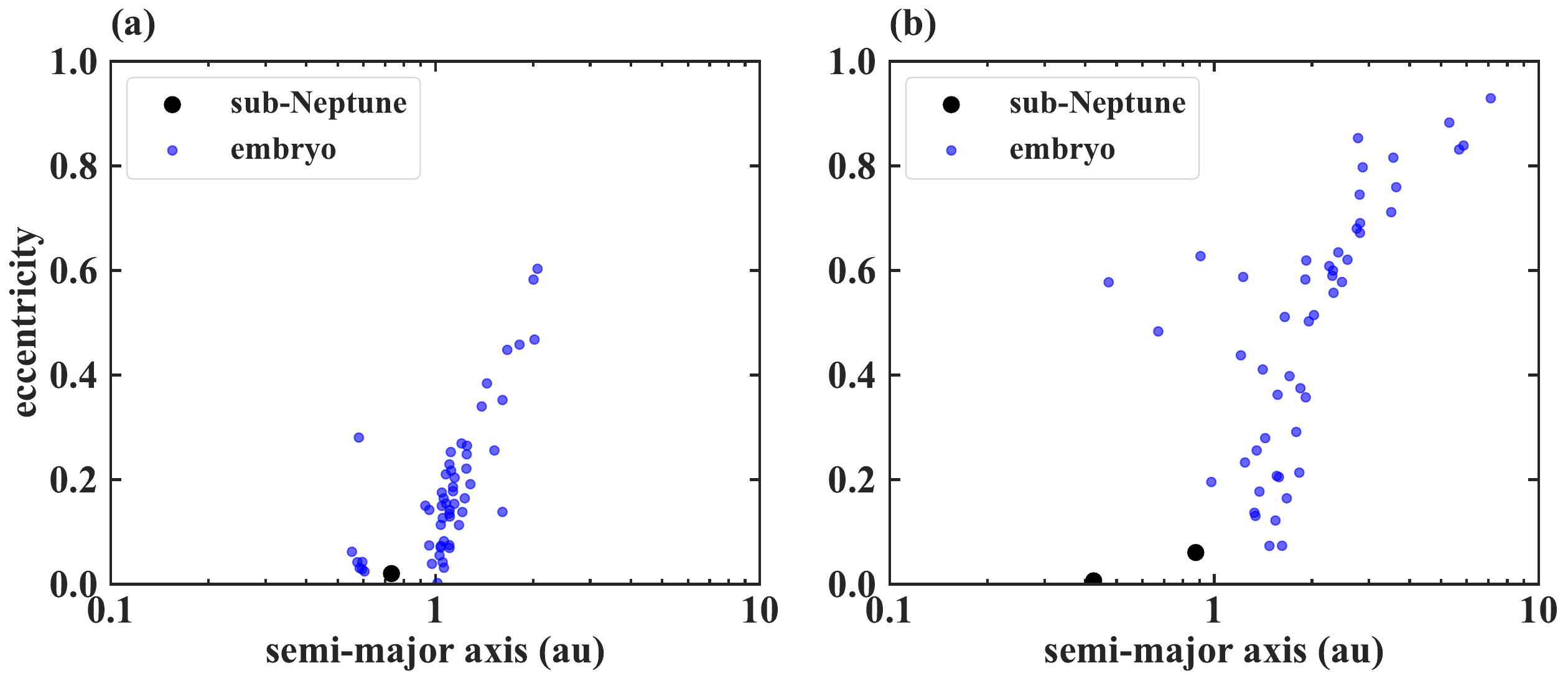}
\caption{Snapshot of the system on semimajor axis and eccentricity at 10\,Myr. The black circles indicate SNs, and the blue circles indicate planetary embryos. Panel (a) represents the result for one SN at initially $a_\mathrm{SN}=0.8$\,au. Panel (b) represents the result for two SNs initially at $a_\mathrm{SN}=0.6$ and $0.8$\,au. 
\label{fig:figA2}}
\end{figure*}

Figure \ref{fig:figA2}(a) and (b) shows the distribution of planetary embryos after 10 $\mathrm{Myr}$ for the cases involving one sub-Neptune and two sub-Neptunes, respectively. From Figure \ref{fig:figA2}(a), we can find that one SN can scatter the planetary embryos, exciting them to relatively high eccentricity, with a maximum eccentricity of about $0.6$. The maximum eccentricity to which a single SN can scatter the embryos is limited because when embryos are scattered to positions farther from the SN, their gravitational interactions become weak. When we put another SN, as shown in Figure \ref{fig:figA2}(b), the innermost SN scatters the embryo into the Hill radius of the second SN, which then further scatters the embryo to a higher eccentricity. The maximum eccentricities that some embryos can reach approach $1$. Therefore, for the initial condition setting, it is reasonable to set the outer embryos to have high eccentricities.

%% For this sample we use BibTeX plus aasjournalv7.bst to generate the

\bibliography{sample701}{}
\bibliographystyle{aasjournalv7}

\end{document}